\documentclass[showpacs, preprintnumbers, amsmath, amssymb,twocolumn]{revtex4}

\usepackage{graphicx}
\usepackage{dcolumn}
\usepackage{bm}
\usepackage{subfigure}

\begin{document}

\preprint{...}

\title{An optical model for an analogy of Parrondo game and designing Brownian ratchets}

\author{Tieyan Si}
\affiliation{Max-Planck-Institute for the Physics of Complex System. Nothnitzer Street 38, D-01187 Dresden, Germany}

\date{\today}

\begin{abstract}
An optical model of classical photons propagating through array of many beam splitters is developed to give a physical analogy of 
Parrondo's game and Parrondo-Harmer-Abbott game. We showed both the two games are reasonable game without so-called game paradox 
and they are essentially the same. We designed the games with long-term memory on loop lattice and history-entangled game. The strong correlation between nearest two rounds 
of game can make the combination of two losing game win, lose or oscillate between win and loss. The periodic potential in Brownian ratchet 
is analogous to a long chain of beam splitters. The coupling between two neighboring potential wells is equivalent to 
two coupled beam splitters. This correspondence may help us to understand the anomalous motion of exceptional Brownian 
particles moving in the opposite direction to the majority. We designed the capital wave for a game by introducing correlations into independent capitals 
instead of sub-games. Playing entangled quantum states in many coupled classical games obey the same rules for 
manipulating quantum states in many body physics.

\end{abstract}


\maketitle

\tableofcontents

\section{Introduction}

It is an old saying from Chinese philosophers 3,000 B.C. that when 
a thing approaches to its extreme point step by step, it will finally transform into the opposite thing\cite{iching}. 
Scientific discoveries stem from different fields to expand this philosophical idea. 
In engineering mechanics, combing two unstable system in the correct way may produce a stable system\cite{allison}. 
Parrondo devised a game to show the combination of two losing games may lead 
to an ultimate wining game\cite{parrondo1996}. This game paradox was used to explain 
the directional motion of molecule motors by 
combining two random Brownian motion according to Parrondo's game paradox\cite{harmer}. In quantum frustrated spin system, 
quantum fluctuation or thermal fluctuation which is supposed to destroy ordering results in long-range magnetic 
ordering\cite{chubukov}, this is termed as order from disorder\cite{villian}.

At first sight, Parrondo's game is counterintuitive. Suppose 
a player has capital $100$ in beginning. He plays two losing game: game A and game B. 
If he only plays game A, the capital will drop. If he only plays game B, the capital drops too. 
If he plays game A many times, the capital drops, he switch to play game B many times, the capital 
continuous to drop, and then he switch to play game A again, and so on. 
One would intuitively expect that the player will finally lose, if the player wins instead, the game may be called a paradox.  
In fact, if there is no any correlation between the nearest two rounds of game, no matter how the player switch between two 
losing games, the final game will lose, it is indeed a paradox if the game wins instead. 
But if there exist correlation between the nearest two 
rounds of losing games or any two rounds of game, the combination of two losing games could be a win, a loss or 
oscillation between win and loss. It depends how the correlation is planted among different games.

Parrondo's game introduced the correlation between the nearest two rounds of game. 
Parrondo, Harmer and Abbott proposed a history-dependent game paradox\cite{parrondo} as an 
improvement of Parrondo's game\cite{parrondo1996}. Later on, the history-dependent game 
paradox was extended to couple two history-dependent games\cite{kay}. The history-dependent game paradox\cite{parrondo} consists of 
two games: game A and game B. The game A is a biased coin which has probability 
$p$ to win and probability $1-p$ to lose. The players of game B are four coins whose strategy at each step 
depends on previous two steps in history\cite{parrondo}. Both Parrondo's game and this history dependent game are 
mathematical issue which study the relationship between the winning or losing probabilities of two coupled games.

An indivisible classical photon scattered by a beam splitter is a perfect analogy of coin. 
If one plays a coin 100 times, it faces up 80 times and faces down 20 times, we say the winning probability 
is $0.8$, and losing probability is $0.2$. Each time the coin faces up, the point increase by $+1$, 
on the contrary, we has $-1$ point upon input capital. The final gain of this game is $(+80-20=+60)$. Therefore this is 
winning game. We let a photon collide with the beam splitter 100 times, it passed the beam splitter 80 times and is reflected 
by the beam splitter for 20 times. The reflection coefficient is $0.2$ and the transmission coefficient is 
$0.8$. Photon is electromagnetic wave. As we know, there is a $e^{i\pi}$ phase shift whenever a wave is reflected. 
Thus the reflected photon naturally carries a negative sign $e^{i\pi}=-1$. When two waves meet 
each other with $\pi$ phase difference, the destructive interference reduce the overlapping intensity to zero. In the language of photon, 
negative photon will annihilate with positive photon. If we let 100 uncorrelated photons collide with the beam splitter, and 
put the reflected photon and transmitted photon together, the number of survived photons is $(+80-20=+60)$. 
The classical photon is not only analogy of a coin, but also a solid physical implementation.

I will use photons passing through beam splitter array to give an almost exact implementation 
of both Parrondo's game and the history-dependent game. 
The difference is Parrondo-Harmer-Abbott game does not 
have the negative sign $-1$ for the lost coins in their mapping matrix. While I will keep the 
$e^{i\pi}$ phase shift for every reflected photon. There are other ways to 
realize similar probability relation, such as special dices, quantum particles passing 
two slits, an so on. A light beam propagating through a network of mirrors(or more accurately beam splitters) 
is a very familiar phenomena for most people. This optical system give us a clear picture to show 
how the probability flow transform from loss to win, and vice verse.

Playing game will produce a time series recording the instantaneous value of output capital. We map 
every time series into an unique path in spatial degree of freedom. One can combine many losing or winning games by drawing 
optical diagrams. A real optical system provide a practical way to test those complex graphical design. 
Parrondo's game only combines two games. 
What I demonstrated in this optical model is a diagrammatic method to combine many games(the number is at least larger than three). 
Every local game may be a loss or win. When many of these games are connected to form a complex network, one has to calculate 
to see the final output. The correlation between neighboring games play the key role in determining the final output.

The coupling between two neighboring beam splitters provide a direct understanding to the directional motion of 
Brownian particles in periodic asymmetric potential. The combination of the two losing games 
in Parrondo's game paradox was used to explain Brownian ratchet\cite{harmer}. In fact, what really matters is the 
strong coupling between neighboring potential wells. Every local potential well behaves like single beam splitter. 
The backward probability is turned into forward probability through two continuous steps of reflection by two neighboring 
potential wells. The optical model suggest that Brownian particle 
in periodic asymmetric potential does not always move in one designed direction, they can oscillate
 back and forth, or move in the opposite direction to the designed one.

The article is organized as follows: 

In section II, we take a classical photon traveling across the array of beam splitters as analogy of 
Parrondo-Harmer-Abbott game. A different transfer matrix from that of Parrondo-Harmer-Abbott game is 
derived by considering the phase shift of reflection beams. When both sub-games lost, 
the final output of this optical game could be a win, a lose or oscillating between a win and a loss.   

In section III, Parrondo game is also expressed into my optical analogy. The modular operation in 
Parrondo game is equivalent to a special beam splitter. The optical diagram has the same structure as 
that of Parrondo-Harmer-Abbott game. 

In section IV, we designed the fractal history tree of a generalized 
history dependent game with longer memory, and extend it to the loop lattice of many beam splitters. We calculated 
the numerical eigenvalues of the transfer matrix for typical cases. When all the 
sub-game lost, the combined game may has oscillating output, or win, or loss.  

In section V, a game with entangled history states is designed. For a symmetric state, two entangled winning game give 
a losing game. For antisymmetric states, the two games liken to lose in order to win the combined game. 
The final output of combined game oscillates between loss and win if only one of the two can win.

In section VI, it shows the optical analogy of Brownian ratchet and how to play a 
quantum wave using classical capitals. A general way of combining 
many quantum sub-games is proposed. It also discussed the relation between quantum many body Hamiltonian and 
the combined game of many sub-games.  

The last section is summary.

\section{An optical analogy of Parrondo-Harmer-Abbott game}

The game A in Parrondo-Harmer-Abbott game paradox is played by a biased coin. The coin has probability $p$ to face up, 
and the probability 
of facing down is $1-p$. If it faces up, the capital increase by one, otherwise the capital decrease by one. 
Suppose we have capital $C$ in the beginning, after many rounds of game A, the winning capital approaches to 
$C\times(+1)\times{p}$. On the losing side, the gain is $C\times(-1)\times{(1-p)}$. 
The final gain of the game is $G=C(2p-1)$. Usually the capital is positive, if the capital is negative, 
we will get a negative value from the winning output, $C\times(+1)\times{p}<0,\;if\;C<0$. While the losing side give a 
positive value, $C\times(-1)\times{(1-p)}>0,\;if\;C<0$. In my optical analogy, 
the negative capital is represented by the photon which is reflected for odd times.

I take an optical beam splitter as analogy of a biased coin. A common Beamsplitters is a half-silvered mirror. Only one part of 
the input beam can pass through the mirror, the other part is reflected. The thickness of the coating aluminum controls the ratio 
between transmitting light beam and reflecting light beam. If the light beam is weak enough so that there is only 
one photon arriving at the mirror within certain time interval. If the photon pass through the mirror, 
the transmitted photon number will increase by one, the transmission probability is $p$. If the photon is reflected, 
then the total number of photons in the reflected beam will increase by one, the reflection probability is $1-p$. 
A photon is also electromagnetic wave, every time it is reflected, a phase shift of $\exp[i\pi]$ will be attached to the photon. Therefore the losing photon 
spontaneously flip a sign. If the time interval between two nearest sequent photons is long enough to erase their correlation, we get a random sequence of game A. 
In the end, the ratio between the total number of transmitted photons and total number of reflected photons is $p/(1-p)$.

\begin{figure}
\begin{center}
\includegraphics[width=0.46\textwidth]{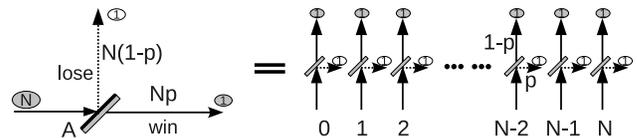}
\caption{\label{gameA} A beam splitter with more reflection than transmission as an implementation 
of game A in Parrando game paradox.The double-faced box with a dark side and a grey side is the beam splitter. 
Many uncorrelated photon passing through one beam splitter is equivalent to many single photon passing through many isolated  
beam splitter.
}
\end{center}
\vspace{-0.5cm}
\end{figure}

Game A can be equivalently implemented by a random sequence of uncorrelated photons. The photon of game A has no memory of history. 
We have two equivalent way to play game A with many photons. 
One way is preparing an identical mirror for each photon, and let the photon pass it only once(Fig. \ref{gameA}). 
The other way is to let many photons pass one mirror(Fig. \ref{gameA}). In the former way, we sum up the number of transmitted photon and 
reflected photon after all the mirrors to get the transmission intensity and reflection intensity. In the most ideal case, 
the value of the intensity of the two beams are the same as measured from the single mirror for many photons. 
Uncorrelated many photon systems is essentially classical light source.

To inject an ideal classical light source into a beam splitter is equivalent to playing game A independently many rounds. 
The number of the rounds of game is the number of photons in the light beam. 
The intensity of the light beam is the capital. The beam splitter is the biased coin. 
The photons transmitted through the mirror are winners, while the reflected photons 
are losers. Suppose the intensity of the input light(or in other words, the initial capital) is $I_0$. 
The total gain of those winning games is ${I_0}p$, while the loss is counted by the reflected photon, $I_0(1-p)$(Fig. \ref{gameA}). 
To make game A as a losing game, the reflected photons must be more than those transmitted. The transmission probability 
is less than a half, $(p=1/2-\epsilon,\;\;\epsilon>0)$.

The game B proposed by Parrondo, Harmer and Abbott\cite{parrondo} can be interpreted as a single Brownian particle 
which determines its strategy for the third step based on the winning or losing states two steps earlier. This game can be  
simulated by computer in a direct way. However it is hard to find ideal correspondence in optical analogy.

We designed an array of beam splitter, and let single classical photon propagate across this array. Here the classical 
photon is real classical object like small metallic ball, even though I called it photon. Every time the classical 
photon pass one beam splitter, it is either reflected or transmitted as one, it never split into two halves. 
We mark the path of transmitted photon by $\textbf{1}$ for it is the winning path. The path of reflected photon is labeled by $\textbf{0}$. 
First, we generate the two-step-earlier historical states. The earliest beam is generated by a single beam splitter, 
it gives one reflection and one transmission. Then we 
place one beam splitter in each way of the output of the earliest beam splitter to generate the 
second earlier beams. Each of the two new beam splitters will generate 
a new pair of transmitted beam and reflected beam. 
Finally we have four historical states recording 
the output beams two steps earlier, 
$\lvert{\textbf{0}_{t-2}\textbf{0}_{t-1}}\rangle$, $\lvert{\textbf{0}_{t-2}\textbf{1}_{t-1}}\rangle$,
 $\lvert{\textbf{1}_{t-2}\textbf{0}_{t-1}}\rangle$, $\lvert{\textbf{1}_{t-2}\textbf{1}_{t-1}}\rangle$.  
If the photon is positive in the beginning, 
the pair of $\lvert{\textbf{0}_{t-2}\textbf{0}_{t-1}}\rangle$ and $\lvert{\textbf{1}_{t-2}\textbf{1}_{t-1}}\rangle$ has positive 
value for the photon has been reflected even times. While the output of $\lvert{\textbf{0}_{t-2}\textbf{1}_{t-1}}\rangle$ and 
 $\lvert{\textbf{1}_{t-2}\textbf{0}_{t-1}}\rangle$ has negative value since the photon is reflected odd times. 
In fact, The two-step history game give out a physical understanding to the so-called game paradox. Suppose both 
the beam splitters at $t-2$ and $t-1$ has a reflection probability $0.8$, the transmission probability is $0.2$.  
The final gain of path  $\lvert{\textbf{0}_{t-2}\textbf{0}_{t-1}}\rangle$ is a positive value, $0.64$. The gain of 
 $\lvert{\textbf{1}_{t-2}\textbf{1}_{t-1}}\rangle$ is $0.04$. The sum of these two positive path, $0.68$, is already larger 
than $0.5$. Therefore, the three losing games of the beam splitters at $t-2$ and $t-1$ lead to a winning results.

\begin{figure}
\begin{center}
\includegraphics[width=0.4\textwidth]{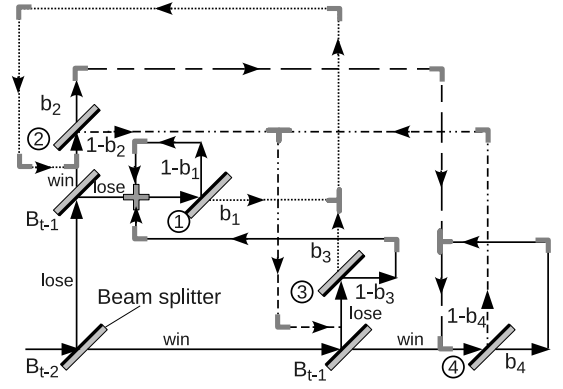}
\caption{\label{mirrorB} The optical flowchart for implementing the game B. The flow lines with arrow guides the flow of photons. 
The closed optical loop demonstrate how to realize the transfer matrix for repeating game B.}
\end{center}
\vspace{-0.5cm}
\end{figure}

To implement the game B proposed by Parrondo, Harmer and Abbott\cite{parrondo}, 
we place the designed beam splitters in game B for each of the four history states. 
Each of the four states only interact with one special beam splitter $m^i_{t}$ at time $t$. Each $m^i_{t}$ 
has a transmission probability of $b_{i}$, and a reflection probability of $(1-b_i)$. 
The games rules are summarized as following(Fig. \ref{mirrorB}),
\begin{eqnarray}\label{gamerule} 
m^1_{t}\lvert{\textbf{0}_{t-2}\textbf{0}_{t-1}}\rangle&=&
b_{1}\lvert\textbf{1}_{t}\rangle+e^{i\pi}(1-b_{1})\lvert\textbf{0}_{t}\rangle,\nonumber\\
m^2_{t}\lvert{\textbf{0}_{t-2}\textbf{1}_{t-1}}\rangle&=&
b_{2}\lvert\textbf{1}_{t}\rangle+e^{i\pi}(1-b_{2})\lvert\textbf{0}_{t}\rangle,\nonumber\\
m^3_{t}\lvert{\textbf{1}_{t-2}\textbf{0}_{t-1}}\rangle&=&
b_{3}\lvert\textbf{1}_{t}\rangle+e^{i\pi}(1-b_{3})\lvert\textbf{0}_{t}\rangle,\nonumber\\
m^4_{t}\lvert{\textbf{1}_{t-2}\textbf{1}_{t-1}}\rangle&=&
b_{4}\lvert\textbf{1}_{t}\rangle+e^{i\pi}(1-b_{4})\lvert\textbf{0}_{t}\rangle.
\end{eqnarray}
Every history states has an unique destination. For instance, if the photon 
is reflected by the beam splitters at time $t-2$ and $t-1$, it will meet beam splitter $m^1_{t}$. The beam splitters at time $t-2$ and $t-1$ are only used for generating 
the initial historical states of game B. After the first two-rounds of game B, the newly generated beams out of $m^i_{t}$ becomes 
the history for the next round. Game B can repeat itself on the four designed beam splitters, $m^i_{t},\;i=1,2,3,4.$ We denote the 
four historical states as a vector,
\begin{eqnarray} 
\lvert\psi\rangle=[\lvert{\textbf{0}_{t-2}\textbf{0}_{t-1}}\rangle,\lvert{\textbf{0}_{t-2}\textbf{1}_{t-1}}\rangle,
\lvert{\textbf{1}_{t-2}\textbf{0}_{t-1}}\rangle,\lvert{\textbf{1}_{t-2}\textbf{1}_{t-1}}\rangle].
\end{eqnarray} 
The transfer matrix map the historical states one step forward in the direction of time, 
\begin{eqnarray} 
\lvert\psi_{t-1,t}\rangle&=&T^{+}\lvert\psi_{t-2,t-1}\rangle.
\end{eqnarray} 
In this optical analogous system, we can directly read out the transfer matrix $T^{+}$ from Fig. \ref{mirrorB}, 
\begin{equation}
T^{+}=\left[
\begin{matrix}
e^{i\pi}\mathring{b}_1 & 0 & e^{i\pi}\mathring{b}_3 & 0 \\
b_1 & 0 & b_3 & 0 \\
0 & e^{i\pi}\mathring{b}_2 & 0 & e^{i\pi}\mathring{b}_4 \\
0 & b_2 & 0 & b_4 \\
\end{matrix}\right],
\end{equation} 
where $\mathring{b}_i$ is the reflection probability, $\mathring{b}_i=1-{b}_i$. 
$T^{+}$ is similar to the matrix given by Parrondo, Harmer and Abbott\cite{parrondo}, but it 
is not mathematically equivalent to that matrix. Here we attached a $e^{i\pi}$ phase shifter to 
every reflection. $T^{+}$ can not be transformed into the matrix in Ref.\cite{parrondo} through elementary 
matrix operation. It has different eigenvalue and different eigenstates. 
We calculated the stationary states $\lvert\psi_{sta}\rangle$ which 
is invariant under the operation of $T^{+}$, $\lvert\psi_{sta}\rangle=T^{+}\lvert\psi_{sta}\rangle,$ it turns out to be 
$\lvert\psi_{sta}\rangle=[0,0,0,0].$ While the stationary solution of the matrix in Parrondo-Harmer-Abbott 
game has complex algebra\cite{parrondo}.

The transfer operator $T^{+}$ can be physically carried out by properly connecting the output with input. The output of 
each beam splitter has fixed destination designed by the game rule. We showed the flow chart 
for implementing the operator $T^{+}$ in the optical network(Fig. \ref{mirrorB}). 
The output beam are fused into designed input beams. The photon is trapped in a cyclic optical network. 
To read out the information of the beam after many rounds, one may disconnect the output from the input. 
If one use one photon to play game B, one must play it many times to determine the distribution of probability out of 
each beam splitter. An equivalent way of playing game B is to input a collection of many uncorrelated photons one time, 
and measure the intensity of output beams.

To get a game paradox in real physical system, we must choose 
the beam splitters with proper reflection and transmission to lose both game A and game B, and make sure 
the combination of game A and game B finally wins. The game A is a single beam splitter, if the 
transmission probability is less than reflection probability, i.e., $p<1/2$, we lose game A. 
As for game B, Ref. \cite{parrondo} gave the value of four winning probability to lose the game B 
in Parrondo-Harmer-Abbott game. However, the matrix in Parrondo-Harmer-Abbott game is not 
mathematically equivalent to the transfer matrix I used in the optical game B. 
So their results is not suitable for determining the transmission probability 
of the four beam splitters here.

I will analyze the optical game B in detail and calculate the distribution of transmission probability for losing this game. 
First we make a fair history states through $B_{t-2}$ and $B_{t-1}$ in Fig. \ref{mirrorB}, both of which 
has transmission probability $p=1/2$. The initial input of $m^4_t$ is $p_{t-2}p_{t-1}=0.25$. The input of 
$m^1_t$ is $(1-p_{t-2})e^{i\pi}(1-p_{t-1})e^{i\pi}=0.25$. The input of  $m^2_t$ and  $m^3_t$ are both 
$(-0.25)$ due to the phase shifter $e^{i\pi}$. At $m^1_t$, the photon has probability $b_1$ to pass and $1-b_1$ 
to reflect. The transmitted photon will go to $m^2_{t+1}$ for it loses at $B_{t-1}$ and wins at $m^1_t$. 
The reflected photon at $m^1_t$ will bent back to  $m^1_{t+1}$. At $t+1$, the number of photon goes to beam splitter $m^i_{t+1}$ are 
$N_{i,t+1},\;\;(i=1,2,3,4)$,
\begin{eqnarray}
N_{1,t+1}&=&0.25e^{i\pi}(1-b_1)-0.25e^{i\pi}(1-b_3),\;\;\;\nonumber\\
N_{2,t+1}&=&0.25b_1-0.25b_3,\nonumber\\
N_{3,t+1}&=&-0.25e^{i\pi}(1-b_2)+0.25e^{i\pi}(1-b_4),\;\;\;\nonumber\\
N_{4,t+1}&=&-0.25b_2+0.25b_4.
\end{eqnarray} 
At time $t+n$, the number of photons to $m^i_{t+n}$ is calculated in the same way by operating the 
transfer matrix $n$ times, 
\begin{eqnarray}\label{matrixn}
\left[\begin{array}{l}
N_{1,t+n}\\
N_{2,t+n}\\
N_{3,t+n}\\
N_{4,t+n}
\end{array}\right]=
\left[
\begin{array}{llll}
e^{i\pi}\mathring{b}_1 & 0 & e^{i\pi}\mathring{b}_3 & 0 \\
b_1 & 0 & b_3 & 0 \\
0 & e^{i\pi}\mathring{b}_2 & 0 & e^{i\pi}\mathring{b}_4 \\
0 & b_2 & 0 & b_4 \\
\end{array}\right]^n
\left[\begin{array}{l}
N_{1,0}\\
N_{2,0}\\
N_{3,0}\\
N_{4,0}
\end{array}\right].
\end{eqnarray}
$[N_{1,0},N_{2,0},N_{3,0},N_{4,0}]$ is the initial number of photon at $t=0$, here  
$\vec{N}_{0}=[0.25,-0.25,-0.25,0.25]$. If the photon is reflected in the latest round, we call it a losing path, otherwise, it is winning path. At time $t+n$, 
the losing paths are $\lvert{\textbf{1}_{t+n-1}\textbf{0}_{t+n}}\rangle$ and $\lvert{\textbf{0}_{t+n-1}\textbf{0}_{t+n}}\rangle$. 
The winning paths are $\lvert{\textbf{0}_{t+n-1}\textbf{1}_{t+n}}\rangle$ and $\lvert{\textbf{1}_{t+n-1}\textbf{1}_{t+n}}\rangle$.
Therefore total number of winning photons are $N_{win}=N_{1,t+n}+N_{4,t+n}$, the losing photons are $N_{lose}=N_{2,t+n}+N_{3,t+n}$. If we want to 
win the game at $t+n$, then ${\lvert}N_{win}{\lvert}>{\lvert}{N_{lose}}{\lvert}$.

We perform similarity transformation upon both sides of Eq. (\ref{matrixn}) 
to diagonalize the transfer matrix without modifying the states vector $\vec{N}$. Each eigenvalue is 
an algebra equation covered many pages.We denotes them as 
$[\lambda_{1},\lambda_{2},\lambda_{3},\lambda_{4}]$. The diagonal transfer matrix reads
 Eq. (\ref{matrixn}) greatly,
\begin{eqnarray}\label{diagn}
\left[\begin{array}{l}
N_{1,t+n}\\
N_{2,t+n}\\
N_{3,t+n}\\
N_{4,t+n}
\end{array}\right]=
\left[
\begin{array}{llll}
\lambda^n_{1} & 0 & 0 & 0 \\
0 & \lambda^n_{2} & 0 & 0 \\
0 & 0 & \lambda^n_{3} & 0 \\
0 & 0 & 0 & \lambda^n_{4} \\
\end{array}\right]
\left[\begin{array}{l}
N_{1,0}\\
N_{2,0}\\
N_{3,0}\\
N_{4,0}
\end{array}\right].
\end{eqnarray}
If the distribution of the transmission probability of four beam splitters satisfy the inequality, 
\begin{eqnarray}\label{inequa}
\lambda^n_{1}N_{1,0}+\lambda^n_{4}N_{4,0}+\lambda^n_{2}N_{2,0}+\lambda^n_{3}N_{3,0}>0,
\end{eqnarray}
the game wins at time $t+n$. This inequality equation is defined within four dimensional cubic, 
$\{b_i\in[0,1],i=1,2,3,4\}$. We divide this four dimensional cubic into 16 sub-cubic by dividing each 
$b_i$ into two halves, $\{b^{win}_i\in(1/2,1],i=1,2,3,4\}$ and $\{b^{lose}_i\in[0,1/2),i=1,2,3,4\}$. One way to 
define a game paradox is the solution of inequity equation (\ref{inequa}) exist in the sub-cubic 
$\{b^{lose}_i\in[0,1/2),i=1,2,3,4\}$. Notice that there are four unknown variables but only one inequality equation, 
the solution space is a three dimensional manifold.

I showed four numerical eigenvalues of the transfer matrix in table \ref{4energy}. 
$\beta_k=\{b_{1},b_{2},b_{3},b_{4}\}$ represents one distribution of the four transmission probability. We calculated 
four special cases: $\beta_1=\{b_{1}=0.2,\;b_{2}=0.2,\;b_{3}=0.2,\;b_{4}=0.2\}$, 
$\beta_2=\{0.8,0.8,0.8,0.8\}$, $\beta_3=\{0.2,0.2,0.8,0.8\}$,  $\beta_4=\{0.8,0.8,0.2,0.2\}$. The corresponding eigenvalue of 
the transfer matrix is shown in table \ref{4energy}. $\beta_1$ represents four losing beam splitters, the corresponding eigenvalue 
$\lambda_1$ is negative, $\lambda_2$, $\lambda_3$ and $\lambda_4$ are almost zero. 
If we play game B odd times according to $\beta_1$, the game 
loses, while play it even times, the game wins. $\beta_1$ is a solution of the inequality equation. 
$\beta_2$ represents four winning beam splitters. It is also one solution of the inequality equation.
 $\beta_3$ and  $\beta_4$ represents two losing beam splitter and two winning beam splitters. For  $\beta_3$, 
$\lambda_1$ and $\lambda_2$ are imaginary numbers, if we play game even times, the contribution of 
$\lambda_1$ and $\lambda_2$ are negative, while $\lambda^2_3$ and $\lambda^2_4$ are positive. But 
the initial input has negative value for $\lambda_2$ and $\lambda_3$, thus the sum of the four paths are almost 
zero. The number $0.77$ in the table is the approximation of $0.774597$. There exist very subtle difference between different 
eigenvalues. The above is just four examples of many solutions. I also checked numerically many other case, 
there are many solutions  within the losing sub-cubic 
$\{b^{lose}_i\in[0,1/2),i=1,2,3,4\}$ which can reach a winning result.

\begin{table}
\begin{center}
\begin{tabular}{lllllllll}
\hline
 &\vline $\lambda_{1}$ & $\vline \lambda_{2}$ &\vline $\lambda_{3}$ &\vline $\lambda_{4}$ \\
\hline
$\beta_1$ &\vline -0.6  &\vline  $10^{-19}+i10^{-9}$ &\vline $10^{-19}-i10^{-9}$ &\vline 0 \\
\hline
$\beta_2$ &\vline 0.8 &\vline $10^{-8}$  &\vline -$10^{-8}$ &\vline $10^{-17}$ \\
\hline
$\beta_3$ &\vline i0.77 &\vline -i0.77 &\vline -0.77 &\vline 0.77 \\
\hline
$\beta_4$ &\vline 0.77 &\vline -0.77 &\vline $10^{-16}$+i0.77 &\vline $10^{-16}$-i0.77  \\
\hline
\end{tabular}
\end{center}
\caption{\label{4energy} The four eigenvalues of the transfer matrix for different transmission probability distribution: 
$\beta_1=\{b_{1}=0.2,\;b_{2}=0.2,\;b_{3}=0.2,\;b_{4}=0.2\}$, $\beta_2=\{0.8,0.8,0.8,0.8\}$, 
$\beta_3=\{0.2,0.2,0.8,0.8\}$,  $\beta_4=\{0.8,0.8,0.2,0.2\}$.}
\end{table}

In analogy of the combination of two games in Parrondo-Harmer-Abbott game\cite{parrondo}, 
I designed the optical network to combine the optical game A and optical game B here. 
It is almost the same network for implementing game B(Fig. \ref{mirrorB}). The beam splitters 
used for generating historical states for game B are now replaced by the beam splitter of game A. Eight beam splitters 
are added to play one more round of game B. When the single photon propagates through the array of beam splitters, 
it shows exactly repeating game A twice followed by two rounds of game B.

\begin{figure}
\begin{center}
\includegraphics[width=0.43\textwidth]{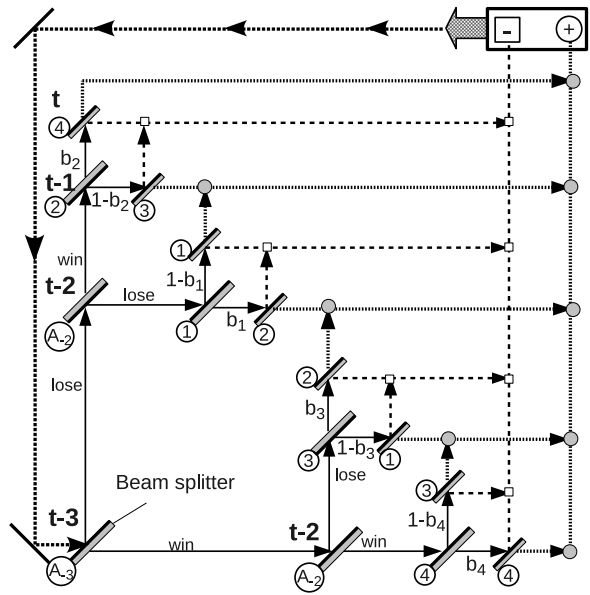}
\caption{\label{mirrorAB} The optical flowchart for playing the combined game in 
Parrondo-Harmer-Abbott history dependent game. The closed loop repeats the game sequence  
$A{\rightarrow} A{\rightarrow}B{\rightarrow}B\cdots$.}
\end{center}
\vspace{-0.5cm}
\end{figure}

First, the photon meet the beam splitter A at time $t-3$(Fig. \ref{mirrorAB}), it has probability $p$ to pass, and probability $(1-p)$ to be reflected. 
In either case, the output photon will be guided to meet another A-type beam splitter at $t-2$. 
There will generate four possible states after the two A-type beam splitters at $t-2$. Until now the photon 
played game A twice to generate the initial historical states for playing game B. At time $t-1$, we place the four 
beam splitters of game B in each way of the four historical paths. It will generate eight possible paths. Following the game 
rule between neighboring historical states in game B, we lead each of the eight possible paths into 
another eight assigned beam splitters at time $t$. So far we played both game A and game B twice.

We input uncorrelated photons one after another into the optical network started 
with A-type beam splitter at time $t-3$, and counts the total number of photons 
running out of each B-type beam splitter at time $t$. The total number of transmitted photon is what we 
win in the combined game. The total number of photons out of the reflection path is the loss. 
To repeat the cyclic game for $n$ steps by single photon,
\begin{eqnarray} 
\overrightarrow{A_{t-3}} \overrightarrow{A_{t-2}}\overrightarrow{B_{t-1}}\overrightarrow{B_{t}}
\overrightarrow{A_{t+1}}\overrightarrow{A_{t+2}}\;\cdots\;\cdots\;\overrightarrow{B_{t+n-1}}\overrightarrow{B_{t+n}},
\end{eqnarray}
we collect the transmitting path out of the B-type beam splitters at time $t$ into one ultimate path, and fuse 
all the reflected paths at time $t$ into another ultimate path(Fig. \ref{mirrorAB}). The difference between
 the two ultimate paths is the net gain. Then we bent the net gain back to the A-type beam splitter at $t-3$. 
The game, $A\rightarrow{A}\rightarrow{B}\rightarrow{B}$, starts all over again
(Fig. \ref{mirrorAB}).

If we do not use single photon to play the game, another equivalent way is to 
input many uncorrelated photons into beam splitter $A_{-3}$ at the same time. 
If the transmitted photons out of the eight beam splitters at $t$ is more than those reflected, the game wins. 
In the language of classical optics, the intensity of transmitted light is stronger than the reflected light. 
In Fig. \ref{mirrorAB}, we use a positive circle to denote the collection of transmitted light beams, 
and a negative square to represent the reflected beams.

This optical game is not equivalent to the mathematical game paradox proposed 
by Parrondo, Harmer and Abbott\cite{parrondo}. 
But it can help us to understand those mathematical issues. For physicist, there is no 
paradox in this optical network. The intensity of reflected light represent 
exactly the losing probability. The transmitted light measures the 
winning probability. Even if there are more reflection than transmission at a local beam splitter, 
as part of the reflected beam can transform into transmitted beam through other beam splitters, 
it is not paradox if the transmitted beam out of the last round 
of beam splitters is stronger than the reflected beam.

In fact, one may design much more complicate optical networks for more optical games. We 
first construct an ultimate optical network that produce 
stronger transmitted beam than reflected beam. 
Then we decompose the total network into small networks, 
and check if the small network generate stronger reflected light than transmitted light. Every beam splitter 
introduce a variable. For a large network with $m$ beam splitters, the solution 
of inequality equation of the final gain is distributed in a $m$-dimensional cubic. As the dimension 
grows, the solution space also expands. One can find many different
ways of decomposing the total network, and each sub-network give more reflection than transmission.

\section{The optical model in analogy of Parrondo's game}

Parrondo's game\cite{parrondo1996} is the combination of two games. Game A is the same as that in the history-dependent 
game, it has probability $p$ to win and probability $1-p$ to lose. Game B first check wether the capital is divisible by 
a number $M$ or not. If the capital can be divided by $M$, it play coin $1$ which has a winning probability $b_1$ and a losing 
probability $1-b_1$, otherwise it play coin 2 which has probability $b_2$ to win and probability $1-b_2$ to lose\cite{parrondo1996}.

\begin{figure}
\begin{center}
\includegraphics[width=0.35\textwidth]{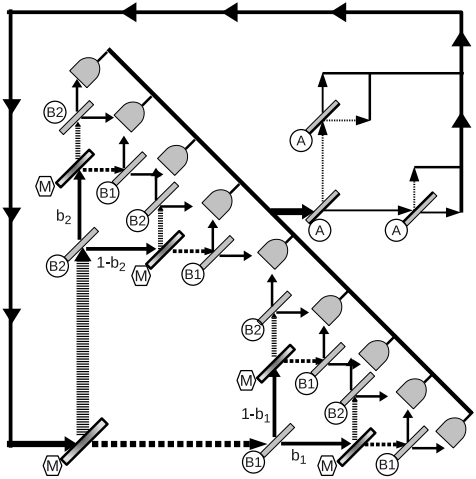}
\caption{\label{divi} The optical flowchart for implementing Parrondo's game. The hexagon with $M$ represents 
the modular beam splitter. The circles with numbers represents the label of beam splitter. Small grey window are the symbol 
of photon detector.}
\end{center}
\vspace{-0.5cm}
\end{figure}

I will design an optical system to theoretically visualize Parrondo's game(Fig. \ref{divi}). 
First we represent every coin by a beam splitter. 
The capital is a collection of many classical photons. Game A is realizable by single beam splitter. Game B needs two 
different beam splitters. The main difficulty of implementing Game B lies in doing the modular algorithm with photons. 
It is a very complex procedure to check if the number of 
photon is divisible by $M$ or not. I just use a hexagonal box to represent a fictional machine which 
can count the number of photons and do the modular computation(Fig. \ref{divi}). I call this machine a modular beam splitter for 
it behaves like an effective beam splitter. For example, if $M=5$, an arbitrary number of 
input photons $N_{input}$ must be one element of the set $\{5k,5k+1,5k+2,5k+3,5k+4\}$. 
We don't know wether $N_{input}$ can be divided by 5 or not. The probability for 
a divisible $N_{input}$ by 5 is $1/5$. $N_{input}$ has a probability 
$4/5$ of being unable to be divided by $5$. So the $N_{input}$ photons has 
probability $4/5$ to meet beam splitter $B_2$, and has probability $1/5$ to meet $B_1$. If we input different number of photons 
into the modular beam splitter for many times, and denote the sum of these input number as $N_t$. The total number of photons 
collected by $B_1$ is $N_t/5$. While $4N_t/5$ of the total number goes to $B_2$. In this sense, the modular beam splitter is 
equivalent to optical beam splitter. However the modular beam splitter does not add phase shifter on its output beams. An optical 
beam splitter add a $e^{i\pi}$ on its reflected beam.

We play this optical game B by bending the output of $B_1$ and  $B_2$ back to the 
initial modular beam splitter, the input capital between the nearest two rounds of game B reads,
\begin{equation}\label{itinerant}
N_{n+1}=N_{n}\left[ \frac{M-1}{M}(2b_2-1)+\frac{1}{M}(2b_1-1)\right]. 
\end{equation}
Suppose the initial capital is $N_0$, then the capital after $k$ rounds of game is, 
\begin{equation}
N_{k}=N_0\left[ \frac{M-1}{M}(2b_2-1)+\frac{1}{M}(2b_1-1)\right]^k. 
\end{equation}
For the special value of the probability parameter $(b_1,b_2)$ and integer $M$ in Ref. \cite{harmer}, 
($M=3$, $b_2=1/10$, $b_1=3/4$), the capital at the $k$th round of game B is $N_k=N_0(-11/30)^k$. If 
k is odd number, $N_k<0$, the game lost. The game wins for an even number of k. 
If both $(b_1,b_2)$ are larger than $1/2$, the net output capital $N_k$ would be constantly positive. For an extreme 
case of $b_1=b_2=1$, i.e., $B_1$ and  $B_2$ has zero probability to lose, then $N_{k}=N_0[+1]^k$. 
For another extreme case, $b_1=b_2=0$, $N_{k}=N_0[-1]^k$ which is positive for $k$ is even. 
This is because if we input a $(-1)$ into a completely losing game, the output would be $+1$. In fact, 
a losing game B has two branches of output. If the computer 
simulation assumes the input capital can only take positive value, the positive branches out of negative input will be eliminated. Only the 
negative branches is left. Then one will see the capital decreases monotonically. The simulation result of game B as showed in 
Parrondo's game is only the negative branch.

Parrondo's game\cite{parrondo1996} is a winning game by periodically repeating game A twice followed by two rounds of game B.  
Fig. \ref{divi} shows the optical flowchart for playing the cyclic game of 
$B\rightarrow{B}\rightarrow{A}\rightarrow{A}\rightarrow{B}\rightarrow{B}\cdots$. The output of game A is 
brought back to the input of game B, it forms a closed loop. 
The game may start from any point in the optical loop. Every transmitted beam out of one modular beam splitter carries 
a number $1/M$. The weight of the reflected beam from modular beam splitter is $(M-1)/M$. The reflected beam out of 
$B_1$ and $B_2$ is attached by an additional phase shifter $e^{i\pi}$ upon their corresponding reflection probability, i.e., 
$e^{i\pi}(1-b_j)$. Playing two rounds of game B from single modular beam splitter give out 16 different output path. The 
final output of each path is calculated by multiplying the weight numbers on all the bonds along this path until it reaches 
the photon collector(the small grey window in Fig. \ref{divi}). The input for beam splitter A at the third round 
is the sum of all the 16 paths. Beam splitter A has reflection probability $p$ and transmission probability $(1-p)e^{i\pi}$. 
After two rounds of game A, the output flows back to the initial modular beam splitter of game B, then the combined 
game starts over again.

We can play this game many times using single classical photon, it will give the same results as we play it once using 
many uncorrelated photon. The game procedure is almost the same as that in the historical game. The main difference is
 the modular beam splitter. Mathematically there is no difference between dividing $N$ by $M$ and dividing $1$ by $M/N$. 
All the analyze about reflection and transmission probability holds for single photon. If we do not divide the 
single photon into many pieces, it is either reflected or transmitted at modular beam splitter. The reflection probability 
is $(M-1)/M$, and transmission probability is $1/M$. The modular beam splitter does not results in $e^{i\pi}$ phase shift 
on its reflection.

Generally speaking, Parrondo's game and Parrondo-Harmer-Abbott history dependent game are the same 
in my optical analogy, except the transfer matrix in my optical model is not mathematically equivalent 
to that in their mathematical game. From a physicist's point of view, both game are reasonable without 
paradox.

\section{Combination of games with long-term memory}

\begin{figure}
\begin{center}
\includegraphics[width=0.4\textwidth]{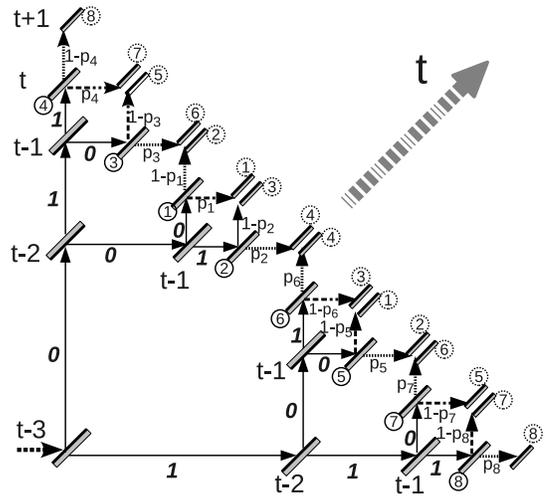}
\caption{\label{tree} The fractal tree array of beam splitters the game with 3-step memory. The $0$ and $1$ on each 
bond are label of losing state and winning state. $t-3$ is the origin point. As $t$ goes to infinity, the 
fractal tree expand larger. The diagram here is the cutoff picture until $t+1$.}
\end{center}
\vspace{-0.5cm}
\end{figure}

We can extend the memory of the game B to $n$-steps of history, $n\geq3$. Every strategy for the next step depends on 
the historical states $n$-steps earlier. So we need at least $n$ beam splitters. Every beam coming out of one 
beam splitter at every step can be clearly marked by reflection beam or transmitted beam. 
We can use new beam splitter to divide every old beam into two beams---a transmitted beam and a reflected beam. 
As long as there is no overlap between any newly generated beams, we construct a fractal tree of beam splitters. 
I will show one extended game with 3-steps of historical memory in the following(Fig. \ref{tree}).

Whenever a photon reach a beam splitter, it is either transmitted or reflected. 
There are eight possible states if the photon pass through three beam splitters one by one. Repeating single photon 
propagation many times is equivalent to propagate many uncorrelated photons once. 
A classical Light beams is a collection of many uncorrelated photons. We count the winners by measuring 
the number of passed photons. The losers are numerated by reflected photons. 
Every path from the beam splitter at $t-3$ to a beam splitter at $t-1$ ends up with 
an unique beam splitter marked by binary code sequence(Fig. \ref{tree}),  
\begin{eqnarray}\label{1-8} 
1_{t}&:=&\lvert{\textbf{0}_{t-3}\textbf{0}_{t-2}\textbf{0}_{t-1}}\rangle,\;\;\;
2_{t}:=\lvert{\textbf{0}_{t-3}\textbf{0}_{t-2}\textbf{1}_{t-1}}\rangle,\nonumber\\
3_{t}&:=&\lvert{\textbf{0}_{t-3}\textbf{1}_{t-2}\textbf{0}_{t-1}}\rangle,\;\;\;
4_{t}:=\lvert{\textbf{0}_{t-3}\textbf{1}_{t-2}\textbf{1}_{t-1}}\rangle,\nonumber\\
5_{t}&:=&\lvert{\textbf{1}_{t-3}\textbf{0}_{t-2}\textbf{0}_{t-1}}\rangle,\;\;\;
6_{t}:=\lvert{\textbf{1}_{t-3}\textbf{0}_{t-2}\textbf{1}_{t-1}}\rangle,\nonumber\\
7_{t}&:=&\lvert{\textbf{1}_{t-3}\textbf{1}_{t-2}\textbf{0}_{t-1}}\rangle,\;\;\;
8_{t}:=\lvert{\textbf{1}_{t-3}\textbf{1}_{t-2}\textbf{1}_{t-1}}\rangle.
\end{eqnarray} 
We introduce a beam splitter operator, $m^j_{t}$, to express the action of the beam splitter ${j}_{t}$ at $t$. 
Beam splitter ${j}_{t}$ let $p_{j}$ of the beam pass, and reflect $1-p_{j}$ of the beam,  
\begin{eqnarray} 
&&m^j_{t}\lvert{\textbf{s}_{t-3}\textbf{s}_{t-2}\textbf{s}_{t-1}}\rangle=
p_{j}\lvert\textbf{1}_{t}\rangle+e^{i\pi}(1-p_{j})\lvert\textbf{0}_{t}\rangle,\nonumber\\
&&j=\textbf{s}_{t-3}\cdot2^2+\textbf{s}_{t-2}\cdot2^1+\textbf{s}_{t-1}\cdot2^0+1.
\end{eqnarray}
Here $j$ corresponds to the number enclosed by a small circle in Fig. \ref{tree}. 
Since every beam splitter corresponds to an unique 3-step historical states. For simplicity, 
we use the sequence of beam splitters at any time $t$ to express the corresponding 3-step historical states before $t$, 
\begin{eqnarray}\label{psit} 
\psi_{t}=\left[\;1_{t},\;2_{t},\;3_{t},\;4_{t},\;
5_{t},\;6_{t},\;7_{t},\;8_{t}\;\right]^{T}.
\end{eqnarray} 
To replay this 3-step history-dependent game, the beam coming out of every beam splitter at time $t$ 
must be guided back into the correct beam splitter at $t+1$. 
The mapping operator from state $\psi_{t}$ to state $\psi_{t+1}$ is
\begin{equation}
T_{_{3}}^{+}=\left[
\begin{matrix}
p_1 & 0 & 0 & 0 & \mathring{p}_5 & 0 & 0 & 0 \\
\mathring{p}_1 & 0 & 0 & 0 & p_5 & 0 & 0 & 0 \\
0 & \mathring{p}_2 & 0 & 0 & 0 & \mathring{p}_6 & 0 & 0 \\
0 & p_2 & 0 & 0 & 0 & p_6 & 0 & 0 \\
0 & 0 & \mathring{p}_3 & 0 & 0 & 0 & \mathring{p}_7 & 0 \\
0 & 0 & p_3 & 0 & 0 & 0  & p_7 & 0 \\
0 & 0 & 0 & p_4 & 0 & 0 & 0 & \mathring{p}_8 \\
0 & 0 & 0 & \mathring{p}_4 & 0 & 0 & 0 & p_8 \\
\end{matrix}\right],
\end{equation}
where $\mathring{p}_i$ is the reflection probability attached by a phase shifter, $\mathring{p}_i=e^{i\pi}(1-{p}_i)$. 
$T_{_{3}}^{+}$ is the transfer matrix between the nearest neighboring time point. 
The states after $n$ rounds of game is reached by operating $T_{_{3}}^{+}$ on the initial 
states $n$ times,   
\begin{eqnarray}\label{T3n} 
\lvert{\textbf{s}_{t+n-3}\textbf{s}_{t+n-2}\textbf{s}_{t+n-1}}\rangle=(T_{_{3}}^{+})^n
\lvert{\textbf{s}_{t-3}\textbf{s}_{t-2}\textbf{s}_{t-1}}\rangle.
\end{eqnarray} 
$\lvert{\textbf{s}_{t-3}\textbf{s}_{t-2}\textbf{s}_{t-1}}\rangle$ represents the initial input generated by 
three historical beam splitters. Following the fractal-tree in Fig. \ref{tree}, every path must 
pass three beam splitters, $\{m_{t-3}, m_{t-2}, m_{t-1}\}$(here we use time index to label the same type of 
beam splitter for simplicity), to reach the beam splitter at time $t$. There are one $m_{t-3}$ beam splitter,
two $m_{t-2}$ and four $m_{t-1}$ beam splitters. The initial intensity of the beam can be read out by direct inspection 
of Fig. \ref{tree},  
\begin{eqnarray}\label{P1-8} 
P_{1}&=&{\mathring{p}_{_{t-3}}\mathring{p}_{_{t-2}}\mathring{p}_{_{t-1}}},\;\;\;
P_{2}=\mathring{p}_{_{t-3}}\mathring{p}_{_{t-2}}p_{_{t-1}},\nonumber\\
P_{3}&=&{\mathring{p}_{_{t-3}}p_{_{t-2}}\mathring{p}_{_{t-1}}},\;\;\;
P_{4}={\mathring{p}_{_{t-3}}p_{_{t-2}}p_{_{t-1}}},\nonumber\\
P_{5}&=&{p_{_{t-3}}\mathring{p}_{_{t-2}·}\mathring{p}_{_{t-1}}},\;\;\;
P_{6}={p_{_{t-3}}\mathring{p}_{_{t-2}}p_{_{t-1}}},\nonumber\\
P_{7}&=&{p_{_{t-3}}p_{_{t-2}}\mathring{p}_{_{t-1}}},\;\;\;
P_{8}={p_{_{t-3}}p_{_{t-2}}p_{_{t-1}}}.
\end{eqnarray}
To play a fair game, the initial capital of eight paths, i.e., the initial number of photons in eight paths, must be the same. 
Thus we assign the same transmission probability to the three historical beam splitter, i.e., 
$(P_{t-3}=P_{t-2}=P_{t-1}=1/2)$.

If we investigate the states of game after finite number of rounds, it is convenient to diagonalize the 
transfer matrix $T_{_{3}}^{+}$ by performing similarity transformation upon Eq. (\ref{T3n}). Then 
$(T_{_{3}}^{+})^n$ is expressed by the eight eigenvalues $\lambda^n_{i},(i=1,2,3,\cdots,8)$, 
\begin{eqnarray}
(T_{_{3}}^{+})^n=Diag[\lambda^n_{1},\lambda^n_{2},\lambda^n_{3},\lambda^n_{4},
\lambda^n_{5},\lambda^n_{6},\lambda^n_{7},\lambda^n_{8}].
\end{eqnarray}
I calculated some numerical value of these eigenvalues for five special cases(\ref{probability}): 
(1) $\beta_1$, eight beam splitters all lose; 
(2) $\beta_2$, eight beam splitters all win; 
(3) $\beta_3$, four beam splitters lose, the other four win; 
(4) $\beta_4$, six beam splitters lose, the other two win; 
(5) $\beta_5$, two beam splitters lose, the other six win. The corresponding eigenvalue are shown in 
table \ref{energy}. The zeroes in the box are approximation of $10^{-10}\sim10^{-19}$. The eigenvalue for the five special cases 
all includes complex number, and neighboring complex eigenvalue are conjugate. 
$\lambda_{7}$ and $\lambda_{8}$ are approximately zero. We study the all-losing case $\beta_1$ as an example. The imaginary part of 
are $\lambda_{1}$ and $\lambda_{1}$ are $88$ times larger than their corresponding real part. We 
take their real part as zero. Then we play the game 2 times with $\beta_1$, the total gain is calculated by 
$G=\sum^8_{i=1} P_{i}\lambda^2_{i}=0.05355$, so the final game wins even if all the eight beam splitters loses.

\begin{table}
 \begin{center}
\begin{tabular}{lllllllll}
\hline
 & $p_{1}$ & $p_{2}$ &$p_{3}$ & $p_{4}$ & $p_{5}$ &$p_{6}$ & $p_{7}$ &$p_{8}$\\
\hline
$\beta_1$ \vline& 0.2 & 0.2 & 0.2 & 0.2 & 0.2 & 0.2 & 0.2 & 0.2\\
$\beta_2$ \vline& 0.8 & 0.8 & 0.8 & 0.8 & 0.8 & 0.8 & 0.8 & 0.8\\
$\beta_3$ \vline& 0.2 & 0.2 & 0.2 & 0.2 & 0.8 & 0.8 & 0.8 & 0.8\\
$\beta_4$ \vline& 0.2 & 0.2 & 0.2 & 0.2 & 0.2 & 0.2 & 0.8 & 0.8\\
$\beta_5$ \vline& 0.8 & 0.8 & 0.8 & 0.8 & 0.8 & 0.8 & 0.2 & 0.2\\
\hline
\end{tabular}
\end{center}
\caption{\label{probability}The five special distribution of the eight beam splitter's transmission probability. }
\end{table}

\begin{table}
\begin{center}
\begin{tabular}{lllllllll}
\hline
 &\vline $\lambda_{1}$ & $\vline \lambda_{2}$ &\vline $\lambda_{3}$ &\vline $\lambda_{4}$ &\vline $\lambda_{5}$ &\vline $\lambda_{6}$ &\vline $\lambda_{7}$ &\vline $\lambda_{8}$\\
\hline
$\beta_1$ &\vline -0.01  &\vline -0.01, &\vline 0.74 &\vline -0.6 &\vline 0.57 &\vline -0.29 &\vline 0 &\vline 0\\
\;\;\;\textit{Im}\;&\vline i 0.88 &\vline -i0.88 &\vline 0 &\vline 0 &\vline 0 &\vline 0 &\vline 0 &\vline 0\\
\hline
$\beta_2$ &\vline -0.42 &\vline -0.42  &\vline 0.83 &\vline 0.83 &\vline 0.6 &\vline 0.19 &\vline 0 &\vline 0\\
\;\;\;\textit{Im}\;&\vline i 0.76 &\vline-i0.76&\vline i 0.02 &\vline -i 0.02 &\vline 0 &\vline 0 &\vline 0 &\vline 0\\
\hline
$\beta_3$ &\vline 0.94 &\vline 0.86 &\vline -0.33 &\vline 0.33 &\vline -0.07 &\vline -0.07 &\vline 0 &\vline 0\\
\;\;\;\textit{Im}\;&\vline  0 &\vline 0 &\vline i 0.71 &\vline -i 0.71 &\vline i0.33 &\vline-i0.33 &\vline 0 &\vline 0\\
\hline
$\beta_4$ &\vline 0.9 &\vline -0.02 &\vline -0.02 &\vline 0.76 &\vline -0.63 &\vline 0 &\vline 0 &\vline 0\\
\;\;\;\textit{Im}\;&\vline  0 &\vline -i0.85 &\vline i 0.85 &\vline 0 &\vline 0 &\vline 0 &\vline 0 &\vline 0\\
\hline
$\beta_5$ &\vline -0.48 &\vline -0.48 &\vline 0.59 &\vline 0.59 &\vline 0.77 &\vline 0 &\vline 0 &\vline 0\\
\;\;\;\textit{Im}\;&\vline  i0.66 &\vline -i0.66 &\vline i 0.5 &\vline 0 &\vline 0 &\vline 0 &\vline 0 &\vline 0\\
\hline
\end{tabular}
\end{center}
\caption{\label{energy} The eight corresponding eigenvalues with respect to 
the transmission probability distribution in table \ref{probability}.
The zeroes in small boxes are approximation of $10^{-10}\sim10^{-19}$. $Im$ indicate the imaginary part of the eigenvalue.}
\end{table}

\begin{figure}
\begin{center}
\includegraphics[width=0.4\textwidth]{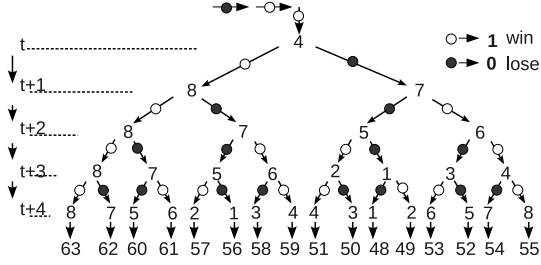}
\caption{\label{fractal} The decimal labeling of all the paths starting from beam splitter 4 until time $t+4$. The small black disc 
indicates the reflection beam. The small white disc represent the transmission beam. The numbers within $1\sim8$ after each black or white disc 
are the label of the beam splitter in game B.}
\end{center}
\vspace{-0.5cm}
\end{figure}

A more geometric way of visualizing this game is to draw a fractal tree of history. Every bond on the tree has a number representing reflection 
or transmission. We follow each path from the origin point to the end, and 
multiply all the numbers on the bond along this path, then we get the gain of this path. The sum of
all different paths give the total gain. For example, there are many future paths starting with $ 4_{t}$ at time $t$. These future paths are given by 
the transfer matrix $T_{_{3}}^{+}$. In Fig. \ref{fractal}, we showed the fractal 
tree of the paths started from  $ 4_{t}$ until $t+4$. Any path is marked by an unique 
binary code sequence. Every three nearest neighboring binary code indicates one historical states. For instance, 
the binary sequence 
\begin{eqnarray}\label{0101}
\textbf{0}_{t-3}\textbf{1}_{t-2}\textbf{1}_{t-1}\textbf{1}_{t}\textbf{0}_{t+1}\textbf{0}_{t+2}\textbf{1}_{t+3}
\end{eqnarray} 
is equivalent to a sequences of 3-step states, 
\begin{eqnarray}
&& \lvert{\textbf{0}_{t-3},\textbf{1}_{t-2},\textbf{1}_{t-1}}\rangle
\rightarrow\lvert{\textbf{1}_{t-2},\textbf{1}_{t-1},\textbf{1}_{t}}\rangle
\rightarrow\lvert{\textbf{1}_{t-1},\textbf{1}_{t},\textbf{0}_{t+1}}\rangle\nonumber\\
& &\rightarrow\lvert{\textbf{1}_{t},\textbf{0}_{t+1},\textbf{0}_{t+2}}\rangle\rightarrow \lvert{\textbf{0}_{t+1},\textbf{0}_{t+2},\textbf{1}_{t+3}}\rangle.
\end{eqnarray} 
In mind of the unique decimal labeling of the beam splitter after the eight historical states, Eq. (\ref{1-8}), we 
get the sequence of beam splitters with different transmission and reflection, 
\begin{eqnarray}\label{48}
 4_{t}\rightarrow8_{t+1}\rightarrow7_{t+2}
\rightarrow5_{t+3}\rightarrow2_{t+4}.
\end{eqnarray}
Every sequence like the above corresponds to an unique path in the fractal-tree network(Fig. \ref{fractal}). 
This path is labeled by $57$ (Fig. \ref{fractal}) which is the decimal mapping of the sequence of binary code Eq. (\ref{0101}), 
\begin{equation} 
0\cdot2^6+1\cdot2^5+1\cdot2^4+1\cdot2^3+0\cdot2^2+0\cdot2^1+1\cdot2^0.
\end{equation}

For a given time $t'$, the path ended up with $\textbf{1}_{t'}$ is winning path. The paths ended up with
$\textbf{0}_{t'}$ are losing paths. The gain(loss) of winning(losing) path 
is calculated by the scalar product of the probability at every beam splitter along the path. We calculate 
the profit of the path sequence Eq. (\ref{0101}) until $t+3$ as an example. 
We denote the number of photons of at beam splitter $4_{t}$ at time $t$ as $N_{0}$. $N_{0}$ is the ultimate result 
of the historical path before $t$. Following the path $\textbf{1}_{t}\textbf{0}_{t+1}\textbf{0}_{t+2}\textbf{1}_{t+3}$, 
$N_{0}{p}_4$ of the $N_{0}$ photons will pass beam splitter $4_{t}$. 
So the gain of this path at time $t$ is $G_{t}=N_{0}{p}_4$. At $t+1$, 
the photons will be reflected by the beam splitter $8_{t+1}$. The number of reflected photons at $t+1$ is 
$G_{t+1}=G_{t}\mathring{p}_8=N_{0}{p}_{4}\mathring{p}_8$, where $\mathring{p}_8=1-{p}_8$ is the 
reflection probability of $8_{t+1}$. 
Following the same calculation, the photons will be reflected at $7_{t+2}$ 
and transmitted through $5_{t+3}$. The final gain of the path 
$4_{t}8_{t+1}7_{t+2}5_{t+3}$ is
\begin{equation} 
G^{(57)}_{t+3}=N_{0}{p}_{4}\mathring{p}_8\mathring{p}_7{p}_5.
\end{equation} 
This is only the gain of path $57$ until time $t+3$. There are $2^7=128$ independent paths at $t+3$, part of which is shown 
in Fig. \ref{fractal}. To find out wether the total game is winning or losing until time $t+3$, 
we must calculate the profit of all the paths, and compare the ultimate number of transmitted photon and 
the number of reflected photons. If the transmitted photon is more than the reflected ones, 
the game wins, otherwise the game loses.

The total number of output paths grows following $2^t$ as time goes on. If we investigate the behavior of the game 
in the long run, it is better to use differential equations with continuous time. Every beam splitter at 
time $t$ corresponds to an unique three-beam splitter sequence in the latest history. Thus if one knows 
the label of beam splitter, one can deduce the information about winning or losing in the latest three steps of game in history. 
For instance, if the label of four beam splitters at time $t$ are $\left\{\;{1},\;{3},
\;{5},\;{7}\right\},$ the game along the four paths at time $t-1$ is losing. For the paths ended up 
with beam splitters $\left\{\;{2},\;{4},
\;{6},\;{8}\right\}$ at time $t$, the game at time $t-1$ is winning. We summaries the eight 
beam splitters into a vector in Eq. (\ref{psit}). The transfer operator $T^+_{_{3}}$ maps a vector $\psi_{t}$ 
to a new vector $\psi_{s}(t+1)$ at $t+1$. The new vector of eight beam splitters bear the same meaning as that 
at time $t$, while the time index increased one step further. Within a longer period, we take the time interval as continuous 
variable so that the difference between neighboring states in history can be denoted as a derivative, 
$\psi_{t+1}-\psi_{t}={\partial}_{t}\psi_{t}$. Then we map the equation of transfer operator into a differential equation,   
\begin{eqnarray}\label{psi} 
\frac{\partial}{\partial{t}}\psi_{t}&=&[{T}_{_{3}}^{+}-I]\psi_{t}.
\end{eqnarray} 
where $[{T}_{_{3}}^{+}-I]$ play the role of Hamiltonian. $I$ is the identity matrix. 
A similarity transformation performed upon both sides of this differential equation 
can diagonalize the Hamiltonian operator by keeping the equation invariant. Properly choosing reflection and transmission probability of the eight beam splitter, one can get an invariant vector 
which keep the same value all the time. This static solution is the eigenfunction of equation 
${T}_{_{3}}^{+}\psi_{t}=\psi_{t}$,

We can design similar flowchart to implement the eight dimensional transfer operator as that for the two-step optical game B. 
The transfer operator redistribute the beams among different beam splitters following the dynamic equation (\ref{psi}). The beams 
passing through some beam splitter may get stronger, while others may become weaker. The time dependent 
distribution is the solution of dynamic equation (\ref{psi}). Exactly diagonalizing the matrix $[T_{_{3}}^{+}-I]$, 
we derived the eigenvalues of the eight beam splitters, 
$\left\{\;E_{j},\;\;j=1,2,3,\cdots,8\;\right\}.$ Each eigenvalue is a heavy algebra equation of 
the eight unknown transmission probability. The solution of the dynamic equation Eq. (\ref{psi}) 
has the following form,  
\begin{eqnarray} 
j(t)=P_{j}\exp[{E_{j}t}],\;\;\; j=1,2,3,..,8.
\end{eqnarray} 
$P_{j}$ is the initial intensity of light beams at time $t=0$. 

We decompose the eigenvalues into real part and imaginary part, $E_{j}=Re_{j}+i\;Im_{j}$. The real part
determines wether the solution diverges or decay to zero as time goes to infinity. 
The imaginary part governs the oscillation of the probability of finding certain beam splitter at time $t$. 
The fractal tree game in Fig. \ref{tree} shows $\{{1}(t+1),{3}(t+1),{5}(t+1),{7}(t+1)\}$ collect winning beams at time $t+1$, 
while $\{{2}(t+1),{4}(t+1),{6}(t+1),{8}(t+1)\}$ absorb the losing beams. Thus the total gain of the game at time $t$ 
is the sum of four winning paths--- $\{{1},{3},{5},{7}\}$, 
\begin{eqnarray} 
G(t)&=&\sum_{j}P_{j}\exp[{E_{j}t}],\;\;\; j=1,3,5,7.
\end{eqnarray} 
The total loss until time $t$ is 
\begin{eqnarray} 
L(t)&=&\sum_{k}P_{k}\exp[{E_{k}t}],\;\;\; k=2,4,6,8.
\end{eqnarray} 
For a winning game, the gain is lager than the loss, ${\lvert}G(t)\lvert>{\lvert}L(t)\lvert$. For a losing game, 
${\lvert}G(t)\lvert<{\lvert}L(t)\lvert$. The inequality equation 
only provide one constraint on the eight unknown variables, i.e., 
$G(t):=G(t,p_{1},p_{2},p_{3},p_{4},p_{5},p_{6},p_{7},p_{8})$. Thus there are many 
different ways to choose proper transmission probability for each of the eight beam splitters. For example, we introduce 
some equations by hand,  
\begin{eqnarray} 
Re[{E_{j}}]&\geq&0,\;\;\;Im[{E_{j}}]=0,\;\;\;\; j=1,3,5,7,\nonumber\\
Re[{E_{k}}]&\leq&0,\;\;\;Im[{E_{k}}]=0,\;\;\;\; k=2,4,6,8.
\end{eqnarray} 
These equations makes the winning beams grow stronger, the losing beams gets weaker. 
The zero imaginary part eliminated the oscillation.

The above procedure of designing optical network for game paradox has a straight extension to 
the $n$-step history-dependent game. However only classical game on network can be analyzed by 
the same strategy of analyzing the three-step history-dependent game paradox. By classical game, I mean 
every path has a definite state of wining or losing. If there exist closed path in network, 
two beams are allowed to inject into the same beam splitter from the opposite surface of the mirror, 
then the transmitted beams from one side may fuse into the reflected beam from the other side. In that case, we 
have to draw a double line---one winning line and one losing line---to label the path.

However, the beam splitter in the loop network does not have unique transmission probability which can be determined 
self-consistently by the same historical mapping rule. For example, if we construct a closed square by placing a single beam splitter at $t+1$ to meet the other three 
beam splitters ${3}_{t}\rightarrow{m}_{t-1}\rightarrow{4}_{t}$ in Fig. \ref{tree}, 
it can not simultaneously satisfy the mapping rule of the vertical path and the horizontal path.
The vertical path leads to ${5}_{t}$, the horizontal path leads to ${7}_{t}$. One way to keep the historical mapping rule 
is popping into the third spatial dimension: place ${5}_{t}$ on top of ${7}_{t}$ at the same lattice site of two dimensions.

\begin{figure}[htbp]
\centering 
\par
\begin{center}
$
\begin{array}{c@{\hspace{0.25in}}c}
\vspace{-0.1cm}\hspace{0.0cm}
\includegraphics[width=0.22\textwidth]{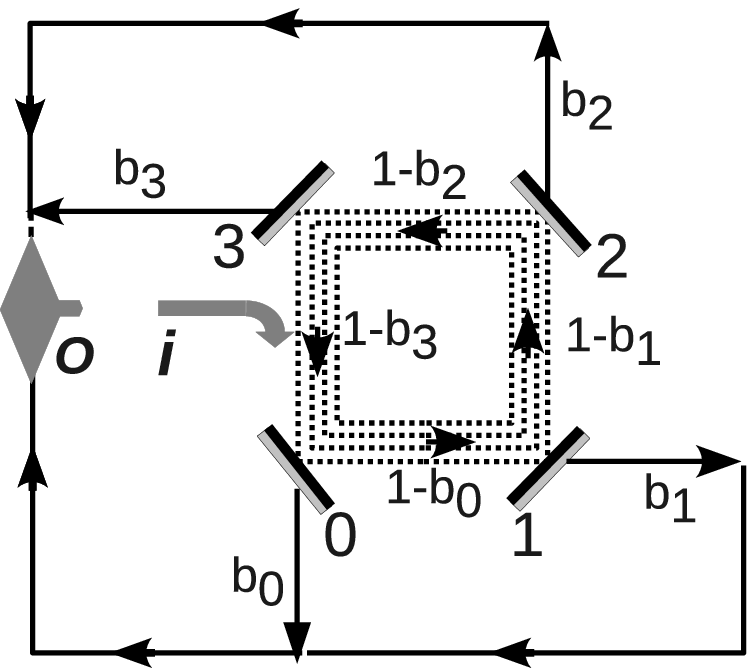}& \includegraphics[width=0.22\textwidth]{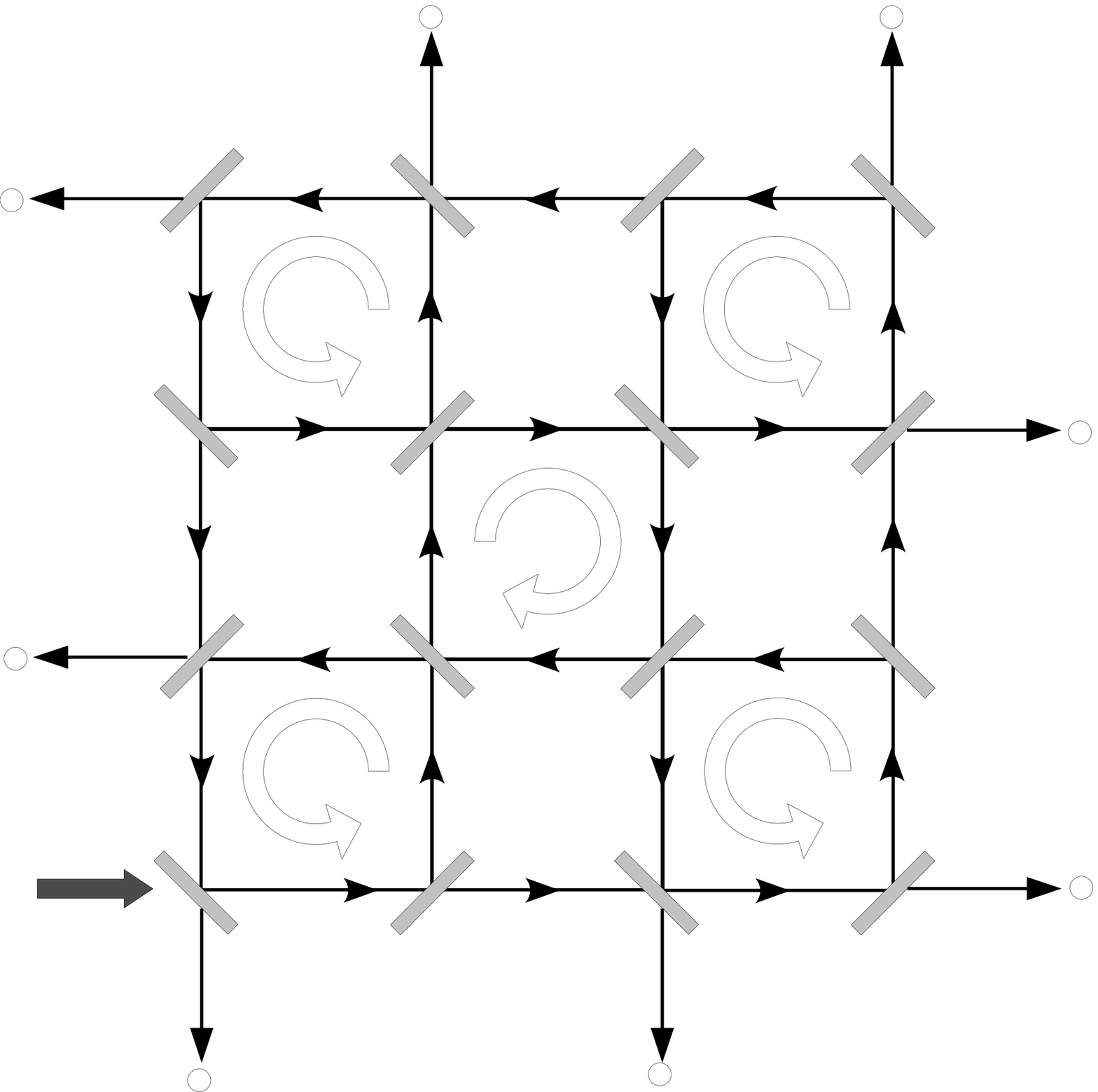}\\
\mbox{(a)} & \mbox{(b)}\\
\end{array}
$
\end{center}
\caption{\label{tra4} (a) A loops game with four beam splitters. All reflection beams are transfer into 
transmission beam. (b) The loop game on square lattice. Each bond is the overlap 
of many transmission and reflection. }  
\vspace{-0.2cm}
\end{figure}

In fact, we can design more general game paradox by abandoning the historical mapping rule. First, we construct a network of 
beam splitters, and define its output channel and input channel. Second, the winning capital is the sum of all the transmitted beams.
The losing is the sum of all reflected channels. Finally, we find the proper transmission probability of each beam splitter so that 
the total winning capital is larger than the losing channel. A special exemplar design is shown in Fig. \ref{tra4} (a) which includes only 
four beam splitters. Each beam splitter $j$ has the probability $b_j$ to win and probability $1-b_j$ to lose, $j=0,1,2,3$. 
We start from game $0$, every time the games loses, 
we switch to next game, $0\rightarrow1\rightarrow2\rightarrow3\rightarrow0\cdots$. The winning beam are collected by the output 
rhombus. Even if each beam splitter has very small transmission probability, $(b_{j}=\epsilon_{j}\ll1,\;\;\;j=0,1,2,3,)$ after 
a enough long time, all the photons will be collected by the output. This is winning game in the end.

The history tree for playing many rounds of the square game with four beam splitters in Fig. \ref{tra4} (a) also have a 
fractal structure. As long as we can distinguish the reflected beam with the transmitted at each round of game, every 
path in history has an unique label. Whenever the beam flows into a beam splitter at some lattice site at certain time, 
the beam splitter must make another copy of itself and split into two different paths 
to separate the overlapped reflection beam and transmission beam. The history tree of 
the game becomes a huge fractal network, especially when 
we extend this square game into a square lattice of many beam splitters.

In mind of the fact we are dealing with the beam of probability flow instead of physical beams so far, we can design a game 
of real light beam on square lattice of many beam splitters like Fig. \ref{tra4} (b). In that case, we can not 
distinguish reflection beam from transmission beam passing through each bond. Firstly, one has to define the output 
beam splitter and input beam splitter along the boundary. The we inject $N$ photons into the input beam splitter, and collect 
the output photon. If the output number of photons is larger than $N/2$, the game is winning, otherwise, the game is lost. 
The lattice structure enclosed by the boundary may be very complex. To create a physical game paradox, we cut the whole 
network into two or more sub-networks in a proper way so that there are less transmitted photons than reflected photons 
through each sub-network. But as a whole network, the transmitted photon is more than those reflected ones.

\section{History-entangled game}

\begin{figure}
\begin{center}
\includegraphics[width=0.45\textwidth]{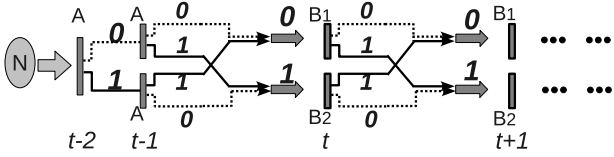}
\caption{\label{quantum} The history-entangled game with two beam splitter in game B. The grey bar represents beam splitters. 
Three A-type beam splitters are used for generating initial historical states.}
\end{center}
\vspace{-0.5cm}
\end{figure}

If the game ask the player to choose its strategy for the next step according to historical record, but 
the player lost some historical information. In this case, the player will select all possible historical 
states to determine its strategy. The final output the game not only depends on historical states, but also 
depends how the player organize those states. I call this kind of game a history-entangled game.

I take a Brownian particle with incomplete memory as player. The particle first travel across 
three identical A-type beam splitter to generate four initial history states(Fig. \ref{quantum}). 
The transmission probability of $A$-type beam splitter is $p_{_{A}}$. Its reflection probability is 
$(1-p_{_{A}})$. If we have capital $N$ in the beginning, then the output capital of the four beams after
 A-beam splitters are(Fig. \ref{quantum})
\begin{eqnarray} 
C_{00}&=&N(1-p_{_{A}})^2,\;\;\;\;
C_{11}=Np_{_{A}}^2,\nonumber\\
C_{01}&=&C_{10}=N(p^2_{_{A}}-p_{_{A}}).
\end{eqnarray}
To make a fair game without dependence on initial condition, we choose exactly $p_{_{A}}=1/2$ for symmetric states, then 
$C_{00}=C_{11}=0.25$, $C_{01}=C_{01}=-0.25$.

When the particle meet beam splitter $B_1$ and $B_2$ at the third step, it only remembers 
wether the past two rounds of game has the same result or different result.  
For instance, if it remembers that it won only once during the latest two rounds of game, but 
it does not remember in which round it wins. If it wins or loses two rounds, 
the only memory kept in its mind is the results of the two rounds are the same. 
The strategy of the particle is to put the output of 
$\lvert{\textbf{0}_{t-2}\textbf{0}_{t-1}}\rangle$ and $\lvert{\textbf{1}_{t-2}\textbf{1}_{t-1}}\rangle$ 
together, and submit them to $B_1$ beam splitter. 
The output of $\lvert{\textbf{0}_{t-2}\textbf{1}_{t-1}}\rangle$ and $\lvert{\textbf{1}_{t-2}\textbf{0}_{t-1}}\rangle$ 
are combined into $B_2$ beam splitter(Fig. \ref{quantum}).

There are two different ways to mix two historical states. One way is adding them up so that switching 
the two beams arrives at the same state. We called it symmetric state, 
\begin{eqnarray} 
\lvert{\phi_{1,t}}\rangle&=&
\lvert{\textbf{0}_{t-2}\textbf{0}_{t-1}}\rangle+\lvert{\textbf{1}_{t-2}\textbf{1}_{t-1}}\rangle,\nonumber\\
\lvert{\phi_{2,t}}\rangle&=&\lvert{\textbf{0}_{t-2}\textbf{1}_{t-1}}\rangle+\lvert{\textbf{1}_{t-2}\textbf{0}_{t-1}}\rangle.
\end{eqnarray}
The other way is subtracting one beam from another 
beam. If we switch the position of the two states, it will generate a $(-1)$ sign on the original state. We call it 
antisymmetric state,
\begin{eqnarray} 
\lvert{\psi_{1,t}}\rangle&=&
\lvert{\textbf{0}_{t-2}\textbf{0}_{t-1}}\rangle-\lvert{\textbf{1}_{t-2}\textbf{1}_{t-1}}\rangle,\nonumber\\
\lvert{\psi_{2,t}}\rangle&=&\lvert{\textbf{0}_{t-2}\textbf{1}_{t-1}}\rangle
-\lvert{\textbf{1}_{t-2}\textbf{0}_{t-1}}\rangle.
\end{eqnarray}
Both the symmetric state and antisymmetric state are similar to the well known Bell state in quantum mechanics. However I just 
use Dirac bracket to denote state for convenience, there is no quantum state here. The symmetric states is comprehensible from the point view of single Brownian particle. 
It is also convenient for computer simulation. Since both 
$\lvert{\textbf{0}_{t-2}\textbf{0}_{t-1}}\rangle$ and $\lvert{\textbf{1}_{t-2}\textbf{1}_{t-1}}\rangle$ will meet
 $B_1$, it is natural to add up their probability. The anti-symmetric states can be simulated by computer, but it 
is hard to explain physically by probability theory of conventional Brownian particle. We keep the antisymmetric 
state as a choice of particle's strategy.

\begin{figure}[htbp]
\centering 
\par
\begin{center}
$
\begin{array}{c@{\hspace{0.15in}}c}
\vspace{-0.1cm}\hspace{0.0cm}
\includegraphics[width=0.23\textwidth]{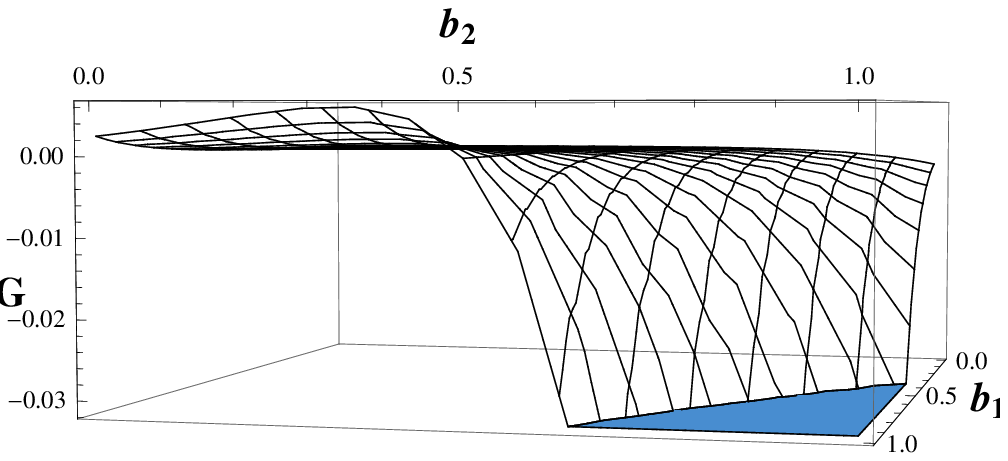}& \includegraphics[width=0.23\textwidth]{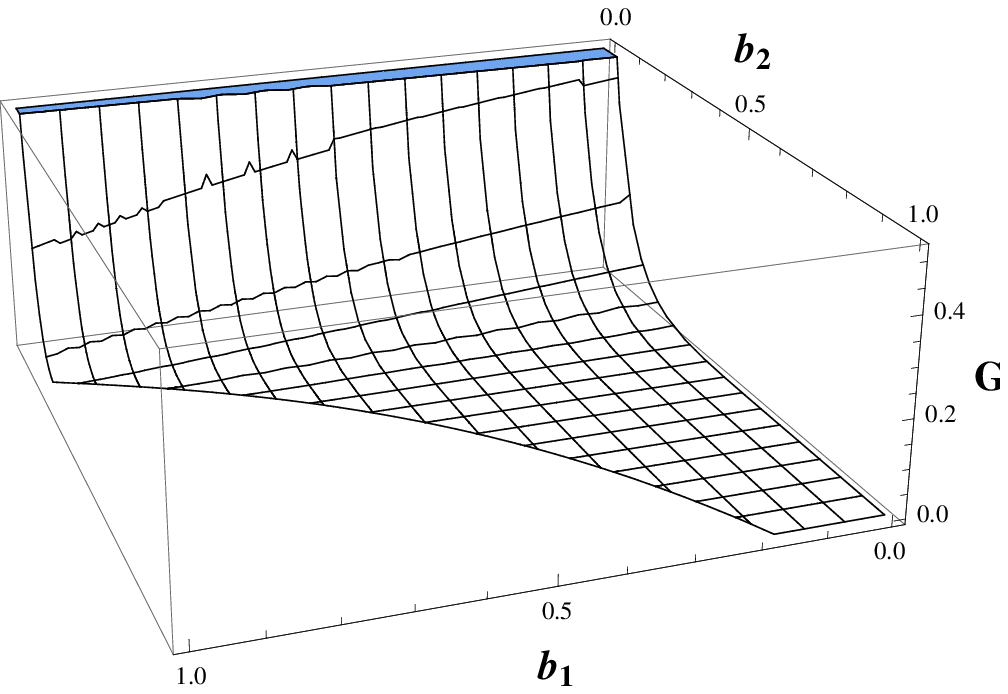}\\
\mbox{(a)} & \mbox{(b)}\\
\end{array}
$
\end{center}
\caption{\label{gboson} (a)  The final gain of symmetric states at t=6. Here we chose $P_0=1$ in the computation 
for convenience. (b) The final gain of antisymmetric states at time t=10.} \vspace{-0.2cm}
\end{figure}

We first study the long term behavior of the symmetric state in by repeating game B. $B_{1}$ take 
$\lvert{\phi_{1,t}}\rangle$, let $b_{1}$ of the input pass 
and reflect $(1-b_{1})$. $\lvert{\phi_{2,t}}\rangle$ will meet beam splitter $B_{2}$. There also generate 
a pair of transmission and reflection at $B_{2}$. When the four output of  $B_{1}$ and  $B_{2}$ at $t$ go to 
the next round game at another pair of $B_{1}$ and $B_{2}$ at $t+1$, the particle began to take the output at $t$ 
as historical states to reorganize 
the latest 2-step history following the same game rule. It view $\lvert{\phi_{1,t}}\rangle$ 
as a losing state $\lvert{\textbf{0}_{t}}\rangle$, and $\lvert{\phi_{2,t}}\rangle$ is winning state 
$\lvert{\textbf{1}_{t}}\rangle$(Fig. \ref{quantum}). Repeating the pair of $B_{1}$ and $B_{2}$ 
will generate a sequence of the pair of states,
\begin{eqnarray} 
\lvert\phi_t\rangle=[\lvert{\phi_{1,t}}\rangle,\lvert{\phi_{2,t}}\rangle]^{T}.
\end{eqnarray} 
The transfer matrix mapping every state vector $\lvert\phi_t\rangle$ one step forward into $\lvert\phi_{t+1}\rangle$ is  
\begin{equation}
T_{\phi}^{+}=\left[
\begin{matrix}
e^{i\pi}(1-{b}_1) & b_2  \\
b_1 & e^{i\pi}(1-{b}_2) \\
\end{matrix}\right].
\end{equation} 
If game B is played so many times that the lifetime of the whole game 
is almost infinite comparing with single round of game, we take state function as 
a continuous distribution, and write the difference equation as the derivative of 
$\lvert{\phi}_{t}\rangle$ with respect to time, 
$\lvert\phi_{t+1}\rangle-\lvert\phi_{t}\rangle={\partial}_{t}\lvert\phi_{t}\rangle$. 
Then the discrete equation, $\lvert{\phi}_{t+1}\rangle=T^{+}\lvert{\phi}_{t}\rangle,$ is equivalently 
mapped into the differential equation, 
${\partial_{t}}\lvert{\phi}\rangle=[T_{\phi}^{+}-I]\lvert{\phi}\rangle.$ As the probability element in 
the transfer matrix have no dependence on time, we can diagonalize the composite matrix 
$[T_{\phi}^{+}-I]$ directly to determine the long term behavior of the states. The two eigenvalues of the 
composite matrix are $E_1=-2, E_2=-2+b_1+b_2$, so the solution of 
this differential equation is.
\begin{eqnarray} 
\lvert{\phi_{1}}\rangle=P_{0}\exp[{-2 t}],\;\;\lvert{\phi_{2}}\rangle=P_{0} \exp[{(-2+b_1+b_2)t}]. 
\end{eqnarray} 
$P_0$ is the initial capital at time $t=0$. The total gain at time $t$ is 
\begin{eqnarray} 
G(t)=P_{0}(2b_1-1)e^{-2t}-P_{0}(2b_2-1)e^{(-2+b_1+b_2)t},
\end{eqnarray}
Here the absolute value of initial input capital is $P_{0}=0.25$. The initial input is determined by 
$C_{00}=C_{11}=0.25$ and $C_{01}=C_{01}=-0.25$. The total gain $G(t)$ at time $t=6$ is plotted in Fig. \ref{gboson} (a).
As time goes to infinity, the total gain approaches to zero, $\lim_{t\rightarrow\infty} G(t)=0$. 
When both game  $B_1$ and $B_2$ wins, the combined game loses. The gain function in Fig. \ref{gboson} (a) drops to negative
in the region $b_2>1/2,\;b_1>1/2$. When $b_2<1/2$, the total gain is positive and increasing as $b_1$ increases
 from $0$ to $1$. Thus if $B_1$ and $B_2$ loses, the game can still win. If beam splitter $B_1$ wins at $b_1=1$, 
while $b_2$ increase from $0$ to $0.32$, the gain is positive and reaches its maximal point at $b_2=0.32$. As $b_2$ continuous 
to increase, the gain drops to zero and finally becomes negative.

The output of antisymmetric states is much more complex than the output of the symmetric states. 
If two Brownian particle has no correlation, playing the two rounds of game twice using one Brownian particle is 
equivalent to playing the two rounds of game only once using two separate Brownian particle. This indeed is true 
for the symmetric states. But to play the antisymmetric states, there must exist a correlation 
between the two Brownian particles. Suppose two Brownian particle meet beam splitters $A_{t-2}$ and  $A_{t-1}$, 
one wins twice, the other lose twice. 
When they get into $B_1$, either the double winning particle or the double losing one must flip a sign to its capital. 
One particle must check the sign of the other one to determine its own sign. If play the 
antisymmetric state using single Brownian particle, there exist a historical correlation between 
the two-winning-step at one time and the two-losing-step at another time. 
If the single Brownian particle wins twice, when it meets $B_1$, it can not decide its own sign. So it 
has to search the history record for the nearest two-losing-step to check its sign. The computer can flip a sign of a number in 
an easy way. In physics, it is a complex design to add a $\exp[i\pi]$ phase shift on a real photon, besides this 
phase shift can only be added after it transmitted through two nearest neighboring
 beam splitters. One may use optical medium to delay the photon. However, if we play many rounds of game, 
the number of optical medium will exponentially increase to infinity.

\begin{figure}[htbp]
\centering 
\par
\begin{center}
$
\begin{array}{c@{\hspace{0.15in}}c}
\vspace{-0.1cm}\hspace{0.0cm}
\includegraphics[width=0.22\textwidth]{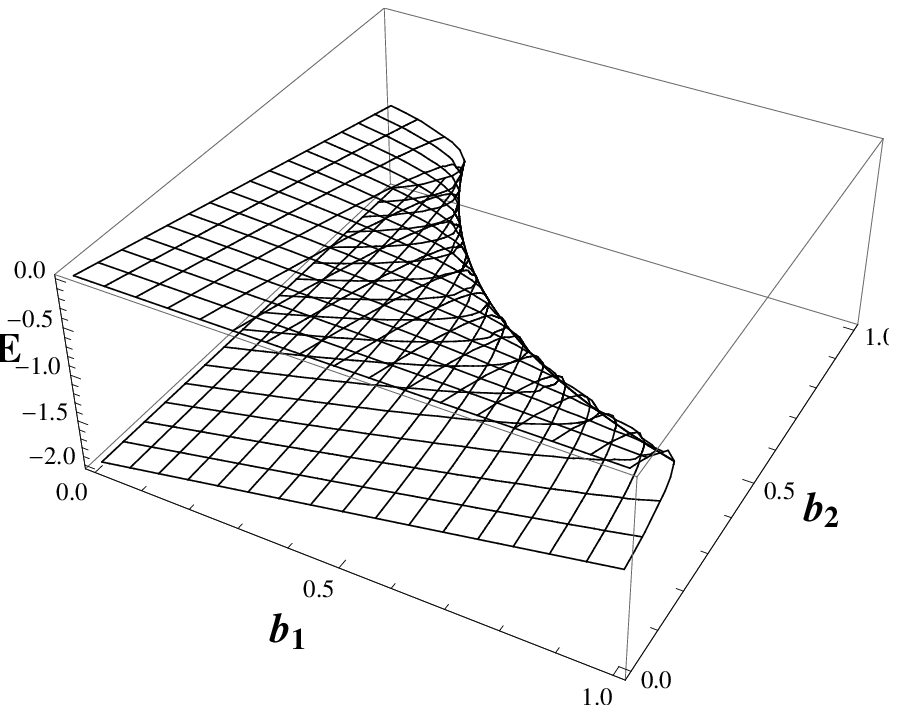}& \includegraphics[width=0.24\textwidth]{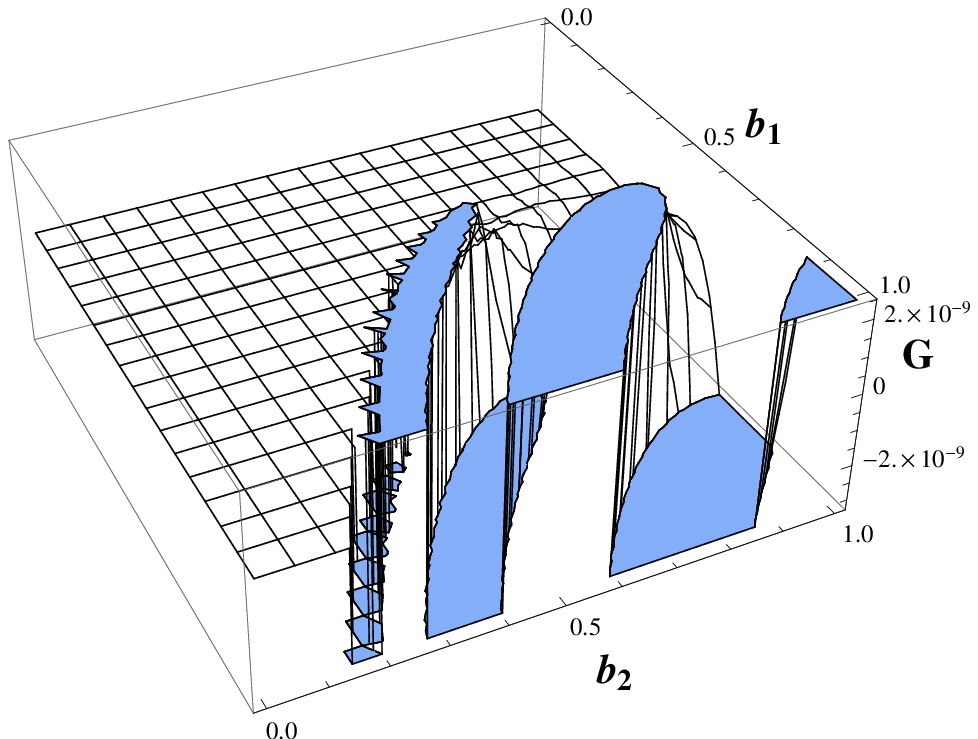}\\
\mbox{(a)} & \mbox{(b)}\\
\end{array}
$
\end{center}
\caption{\label{eigen} (a) The two eigenvalues of the transfer matrix for antisymmetric states,$E_{-}$ and $E_{+}$. 
They are both negative. (b) The final gain of antisymmetric states at t=17 within the oscillating region.} \vspace{-0.2cm}
\end{figure}

The long term behavior of the antisymmetric state is significantly different from symmetric states. 
The game rule of beam splitter $B_1$ and $B_2$ for antisymmetric states is the same as that for symmetric states,
\begin{eqnarray} 
B_{1}\lvert{\psi}\rangle&=&
b_{1}\lvert\textbf{1}_{t}\rangle+e^{i\pi}(1-b_{1})\lvert\textbf{0}_{t}\rangle,\nonumber\\
B_{2}\lvert{\psi}\rangle&=&
b_{2}\lvert\textbf{1}_{t}\rangle+e^{i\pi}(1-b_{2})\lvert\textbf{0}_{t}\rangle.
\end{eqnarray}
The equation of motion for the antisymmetric state vector, 
$\lvert{\psi}\rangle:=[\;\lvert{\psi_{1}}\rangle,\;\lvert{\psi_{2}}\rangle\;]^T,$ is 
${\partial_{t}}\lvert{\psi}\rangle=[T_{\psi}^{+}-I]\lvert{\psi}\rangle.$ The transfer matrix between two nearest 
neighboring historical states is 
\begin{equation}
T_{\psi}^{+}=\left[
\begin{matrix}
e^{i\pi}(1-{b}_1) & -b_2 &  \\
b_1 & -e^{i\pi}(1-{b}_2) &  \\
\end{matrix}\right].
\end{equation} 
$[T_{\psi}^{+}-I]$ may be view as an effective Hamiltonian which has two eigenvalues,
\begin{eqnarray} 
E_{-}&=&\frac{1}{2}(-2-b_2+b_1-\sqrt{\Delta}),\nonumber\\
E_{+}&=&\frac{1}{2}(-2-b_2+b_1+\sqrt{\Delta}).
\end{eqnarray}
where $\Delta=(2+b_2-b_1)^2-8 b_2$ is the gap between two levels. 
Both the two eigenvalues are negative(Fig. \ref{eigen}). So the two antisymmetric states will converge to zero 
when time goes to infinity. This a draw game in the long run. But its output in finite time domain could be winning, losing, 
or oscillating. We analyze the final gain, 
\begin{eqnarray} 
G(t)=(2b_1-1)\lvert{\psi_{1}}\rangle-(2b_2-1)\lvert{\psi_{2}}\rangle,
\end{eqnarray}
from the time-dependent solution of the two antisymmetric states,
\begin{eqnarray} 
\lvert{\psi_{1}}\rangle&=&P_{0}\exp[\frac{1}{2}(-2-b_2+b_1-\sqrt{\Delta})t],\nonumber\\
\lvert{\psi_{2}}\rangle&=&P_{0}\exp[\frac{1}{2}(-2-b_2+b_1+\sqrt{\Delta})t].
\end{eqnarray}
The gain function $G(t)$ has three different branches according to the gap $\Delta$. 

(1) If $b_2=2-2\sqrt{2}\sqrt{ b_1}+ b_1,$ then $\Delta=0$, 
two states become degenerated. In the degenerated states, if  $b_1> b_2,$ the game wins, otherwise the game loses. Notice 
that the results only depends on relative ratio between $b_1$ and $b_2$, even both $b_1$ and $b_2$ loses, i.e., 
$b_1<1/2$ and $b_2<1/2$, the game can still be winning as long as $b_1> b_2$. On the other hand, if both $b_1>1/2$ and $b_2>1/2$,
 the game can also loses if $b_1< b_2$. There also exist others case, such as one of the two game loses, 
the total game loses or wins.  

(2) If $b_2<2-2\sqrt{2}\sqrt{ b_1}+ b_1,$ the gap is positive $\Delta>0$. The game $b_2$ has 
to lose to make a winning combined game. In the winning region, when $b_1$ increases from $0$ to $1$, 
the gain decreases(Fig. \ref{gboson}). Thus $b_1$ also inclined to lose to win the combined game.

(3) If $b_2>2-2\sqrt{2}\sqrt{ b_1}+ b_1,$ the gap is negative, $\Delta<0$.
In this region, the two game can not lose simultaneously. When $b_1$ loses, $b_2$ is always larger than $1/2$. 
If $b_2$ lose, $b_1$ slide into the winning region. There is only one point of draw game, $b_1=b_2=1/2$. However, 
this region allows the two game to win simultaneously, while the final gain of the combined game does not always win. 
$G(t)$ oscillates between winning and losing(Fig. \ref{eigen} (b)). Although the gain 
is not real number but a imaginary number, in some sense, we still call this a game paradox since combining 
two winning game give a losing result.

The different behavior of the final gain between symmetric states and antisymmetric states implies antisymmetric 
states is much more efficient to win the global game by losing every local games. For a history dependent game 
with long term memory, similar antisymmetric state can be constructed following the same strategy. 
The dimension of probability variables will grows higher. There is much freedom for 
one to design various different game paradox.

\section{Optical game as analogy of Brownian ratchets and many body physics}

\begin{figure}
\begin{center}
\includegraphics[width=0.38\textwidth]{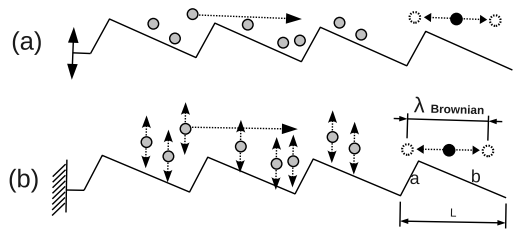}
\caption{\label{rachet}(a) The periodic potential is turned on and off. (b) The periodic potential 
is constantly presented. While the Brownian particles as a whole is in and out of the potential periodically or randomly.}
\end{center}
\vspace{-0.5cm}
\end{figure}

A classical Brownian particle trapped in a periodic potential shows random motion, no matter the local potential well is 
symmetric or asymmetric. If the periodic potential with asymmetric local potential is turned on and off periodically, 
(Fig. \ref{rachet} (a)), the Brownian particle would move in one direction\cite{rousselet}. 
An equivalent way to see this phenomena is to fix the periodic potential but let 
all the Brownian particles together approach to the potential and leave there quickly(Fig. \ref{rachet} (b)). 
This point of view describes the same thing 
by placing the static observer of reference coordinates on the potential well. Thus 
Brownian particles move to the right too.

Every asymmetric potential well is analogous to a beam splitter. When the particle hit the potential well, it tends to stay 
in a lower potential point. If it collides with bond $b$ in Fig. \ref{rachet} (b), moving to the right is the dominant motion. 
When it hits bond $a$, it moves to the left(Fig. \ref{rachet} (b)). Switching the potential on and off in a periodic way in fact is
equivalent to generate a standing wave of Brownian particles upon a static potential. In one dimension,
 Brownian particle has spontaneous random motion along the length of the periodic potential. If we ignore the collision 
between different Brownian particles, one Brownian particle represents one propagating wave. 
As a rough phenomenological description, if a Brownian particle 
can cover an average distance $\lambda_{Brownian}$ during one period of oscillation, $\Delta{T}$, 
we define $\lambda_{Brownian}$ as random wavelength. In one dimension, $\lambda_{Brownian}$ is along the length of potential.  
The one domensional periodic potential is composed of many identical local potential wells which has unit length $L$(Fig. \ref{rachet} (b)).

\begin{figure}
\begin{center}
\includegraphics[width=0.38\textwidth]{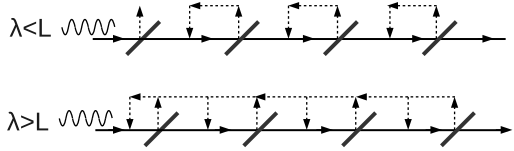}
\caption{\label{wave}(a) The Brownian wave length within one period is smaller than the length of one local potential well. 
(b) The Brownian wave length is larger than the length of an unit of potential well.}
\end{center}
\vspace{-0.5cm}
\end{figure}

If the random wave length $\lambda_{Brownian}$ is smaller than $L$, the propagating wave only collide with one beam splitter, 
or one potential well(Fig. \ref{wave}). The transmission probability is $p=b/(a+b)$, while the reflection probability is 
$1-p=a/(a+b)$. Here the definition of transmission and reflection depends on the direction of input wave, 
one may switch $a$ and $b$ for different cases. If the random wave length $\lambda_{Brownian}$ is larger than $L$, the propagating wave will cover 
two potential wells at the same time. In that case, the wave will pass two beam splitters within one period. We must sum up 
all possible path within two-steps, it has been shown in my optical model of Parrondo-Harmer-Abbott game. But here 
the case is simpler for all the beam splitters are identical. If the random wave length $\lambda_{Brownian}$ is 
so large that it covers many potential wells at the same time,  we must take into account of 
three or more beam splitters at one time. This case meet the history dependent game with long term memory.

\begin{figure}
\begin{center}
\includegraphics[width=0.42\textwidth]{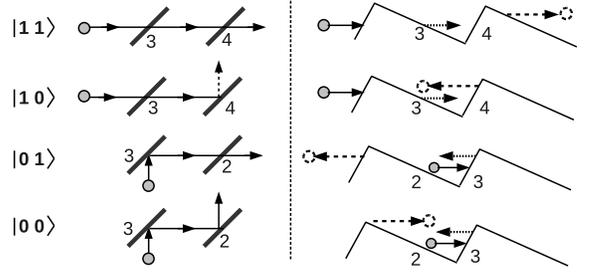}
\caption{\label{four}The four elementary optical diagrams for a particle propagating through two beam splitters. 
The right hand side is the corresponding behavior of a Brownian particle colliding with two neighboring potential wells. }
\end{center}
\vspace{-0.5cm}
\end{figure}

The two or more steps of collision between neighboring potential wells can be mapped into 
a limited number of elementary optical diagrams. I showed the optical diagram of two steps of collision in Fig. \ref{four}.  
The transmission is in the same direction as the incoming velocity. While the reflection is in 
the opposite direction. Every reflection bond is labeled by a number, $-(1-p)=e^{i\pi}(1-p)$. If 
Brownian particle pass two potential wells continuously, we label it as state $\lvert\textbf{11}\rangle$. 
The final output particle moves in the same direction as the input velocity. 
The state $\lvert\textbf{10}\rangle$ denotes that the Brownian particle first pass one potential well 
and then is reflected by the next potential well. The final output velocity of state $\lvert\textbf{10}\rangle$ is 
in the opposite direction of the input velocity. So does the state  $\lvert\textbf{01}\rangle$. For the state 
$\lvert\textbf{00}\rangle$, the Brownian particle is firstly reflected into the opposite direction 
at the first potential well, then it is reflected back to 
the original direction by the second potential well. The final velocity is in the same direction 
as input velocity. With this correspondence between optical diagram and collision within two neighboring 
potential wells, one can calculate the final output velocity of Brownian particles in flashing ratchet. 
Adding up the probability of $\lvert\textbf{00}\rangle$ and $\lvert\textbf{11}\rangle$ gives the positive velocity,
\begin{equation}
V^+=P(\lvert\textbf{00}\rangle)+P(\lvert\textbf{11}\rangle). 
\end{equation} 
The negative velocity is determined by $\lvert\textbf{00}\rangle$ and $\lvert\textbf{11}\rangle$, 
\begin{equation}
V^-=P(\lvert\textbf{01}\rangle)+P(\lvert\textbf{10}\rangle). 
\end{equation} 
If $\lvert V^+\rvert>\lvert V^-\rvert$, the particle moves in the same direction as its original velocity. On the 
contrary case, the particle move in the opposite direction. If $\lvert V^+\rvert=\lvert V^-\rvert$, the particle 
is trapped in a local potential well and moves around randomly. If the length of unit potential well is much 
smaller than the phenomenological random wavelength, 
we must include the three-steps of collision or even more steps. The strategy is the same as I showed above.

\begin{figure}[htbp]
\centering 
\par
\begin{center}
$
\begin{array}{c@{\hspace{0.2in}}c}
\vspace{0.1cm}\hspace{0.0cm}
\includegraphics[width=0.2\textwidth]{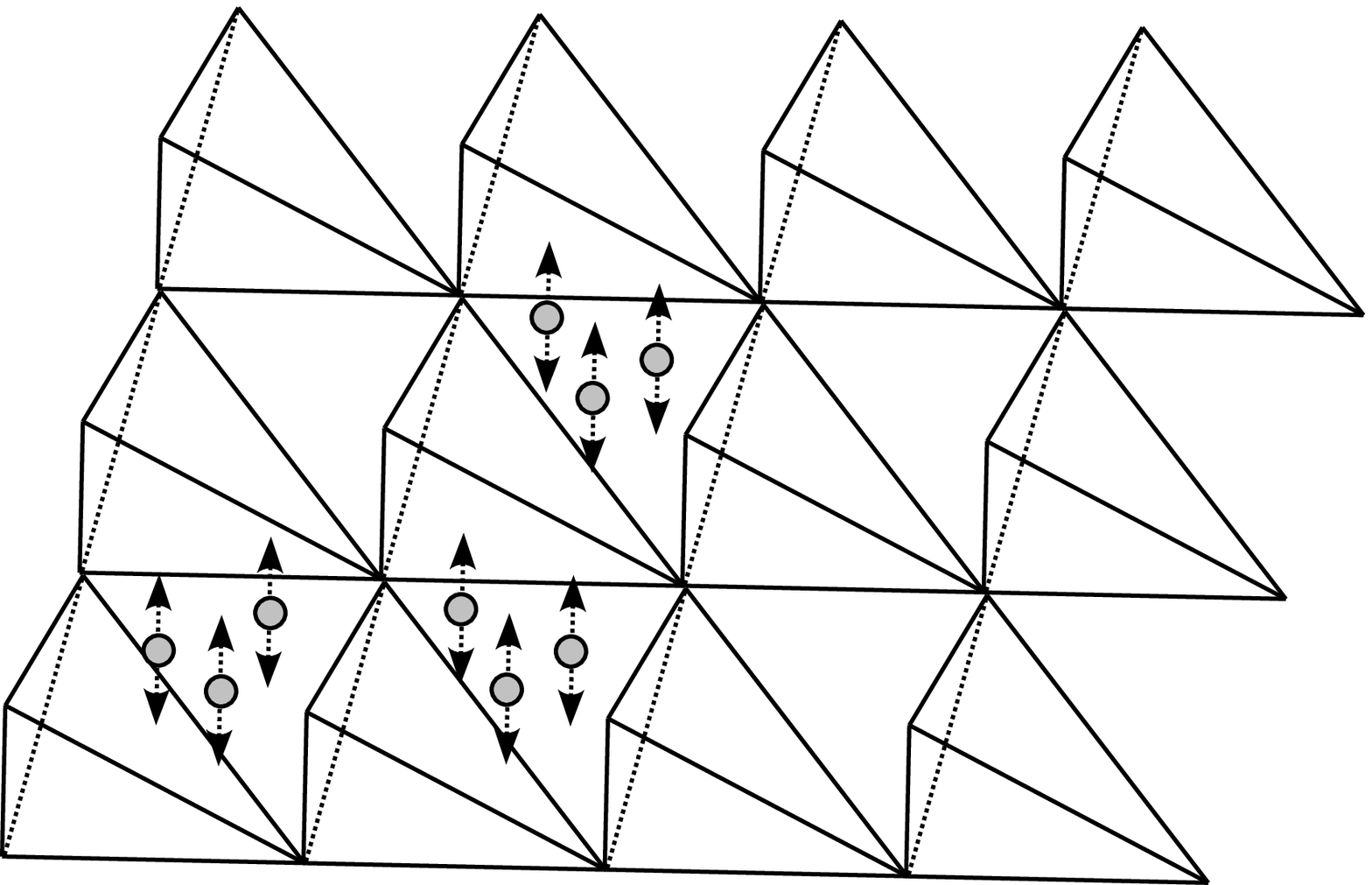}& \includegraphics[width=0.25\textwidth]{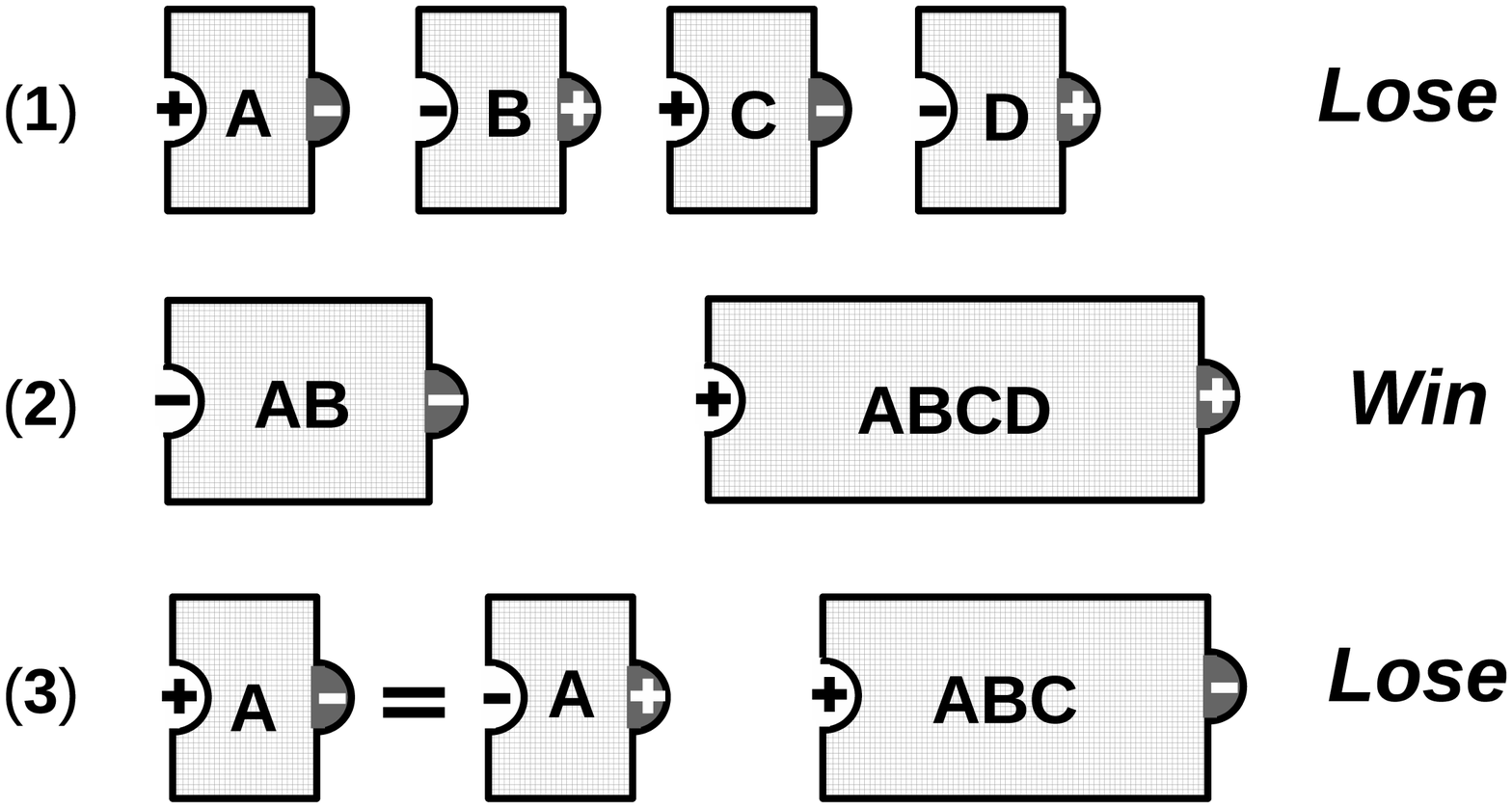}\\
\mbox{(a)} & \mbox{(b)}\\
\end{array}
$
\end{center}
\caption{\label{convert} (a) The asymmetric potential well for Brownian rachet periodically 
distributed in two dimensions. (b) Elementary flipper and its combination. Each box represents a flipper which corresponds to 
one losing game. A losing game is operated by
 odd number of elementary flipper. A winning game can be performed by even number of flipper. } \vspace{-0.2cm}
\end{figure}

If Brownian particles do not collide with each other, the physics of single particle 
is an ideal meanfield approximation of many particle system. 
In fact, the random wave length $\lambda_{Brownian}$ is only well defined for 
a crowd of many Brownian particle. 
When many Brownian particles are trapped in a periodic potential, the collision between different particles 
are modulated to form quasi-regular density wave. When this density wave propagates across the periodic potential, 
it is either reflected or transmitted. Modeling many colliding Brownian particles as density wave is convenient for 
describing and calculating Brownian rachet in two dimensional periodic potential(Fig. \ref{convert} (a)). 
In two dimensional periodic potential, a Brownian particle can 
circumvent potential wells instead of being blocked there. If the Brownian particle can randomly fluctuate around a circle, 
such as sperm or bacteria, the angular velocity is equivalent to an effective magnetic field. In quantum Hall effect 
of electrons in magnetic field, a series of plateaus appear in the Hall resistance measurement. This give us a hint that 
the two dimensional flashing rachet for circling Brownian particle maybe has significant different 
phenomena from one dimension.

When the local density of Brownian particle oscillates periodically under the modulation of external potential, 
it may be called as density wave. It is hard to control exactly the local density, but in a game, 
a player can determine the amount of his capital for each round of game. If we play the capital in 
a proper way, it will generate a good correspondence between Brownian density wave and capital wave of game.

First, we decompose the input capital as the integration of a pair 
of conjugated complex function, $N=\int{\psi^{\ast}\psi}=\langle\psi\lvert\psi\rangle$, $\psi_t$ is 
a wave function whose amplitude is the square root of the total capital, 
\begin{equation}
\psi=\psi_a(t)+i\psi_b(t)=\sqrt{N_t}\exp[{i\Omega{(t)}}],
\end{equation}
The angular velocity $\Omega(t)=\arctan[{\psi_b(t)/\psi_a(t)}]$ is a complicate function of time.  The capital now has internal degree of freedom. Suppose we input
 capital $1000$ points within time interval $T$. But they do not join the game one after another in 
constant speed or once and for all. We divide  those time interval into $n$ pieces, $T=[0,T_1)\cup[T_1,T_2)\cup[T_2,T_3)\cdots\cup[T_{n-1},T_n]$. 
The input capital in each step is $N_{T_{i}}$ which is oscillating within $T$, and $\sum_{i}N_{T_{i}}=1000$. For example, 
we input an cosin wave, $N=\cos[\Omega{(t)}],\;t\in{T}$. Then $N_{T_{i}}=\int^{T_{i}}_{T_{i-1}}\cos[\Omega{(t)}].$ The discrete capitals within $T$ now are not unrelated numbers, they organized 
into waves. This give us the real part of the capital wave $Re[\psi]=\sqrt{N}\cos[\Omega{(t)}]$. The imaginary wave 
of the capital is the delayed or advanced real wave by a phase of $\pi/2$, $Im[\psi]=\sqrt{N}\cos[\Omega{(t)}\pm\pi/2]$. 
The total input capital is 
\begin{equation}
 N=Re[\psi]^2+Im[\psi]^2.
\end{equation}
When a beam splitter received the capital $Re[\psi]^2$ and $Im[\psi]^2$, it take them as independent beams 
and redistribute them separately. Both the real wave and delayed real wave(or imaginary wave) generate 
a transmission wave and reflection wave, we sum the transmission up and denote it as $\psi_{\textbf{1}}$.  
The sum of refection is $\psi_{\textbf{0}}$. The total capital wave if the sum of the two waves, 
$\psi=\psi_{\textbf{1}}+\psi_{\textbf{0}}$. 
We use Dirac bracket to express the winning wave as $\lvert{\textbf{1}}\rangle\rightarrow\psi_{\textbf{1}}$, the 
losing wave is $\lvert{\textbf{0}}\rangle\rightarrow\psi_{\textbf{0}}$, 
$\lvert{\psi}\rangle=\lvert{\textbf{1}}\rangle+\lvert{\textbf{0}}\rangle.$ Then the absolute value of total capital reads
\begin{eqnarray}
N=\langle{\textbf{1}}\lvert{\textbf{1}}\rangle+\langle{\textbf{0}}\lvert{\textbf{0}}\rangle.
\end{eqnarray}
We have assumed there is no overlap between reflection and transmission, thus 
$\langle{\textbf{0}}\lvert{\textbf{1}}\rangle=\langle{\textbf{1}}\lvert{\textbf{0}}\rangle=0$. 
$\langle{\textbf{1}}\lvert{\textbf{1}}\rangle=\rho_{11}$ is transmission probability determined by the beam splitter. 
$\langle{\textbf{0}}\lvert{\textbf{0}}\rangle=\rho_{00}$ is the reflection probability.

If the input angle between capital wave and beam splitter is so small that reflection overlaps transmission, 
the transformation between reflection and transmission is not zero, i.e., 
$\langle{\textbf{0}}\lvert{\textbf{1}}\rangle\neq\langle{\textbf{1}}\lvert{\textbf{0}}\rangle\neq0$. In that case, the total 
transmission is $N_{\textbf{1}}=\langle{\textbf{1}}\lvert{\textbf{1}}\rangle+\langle{\textbf{1}}\lvert{\textbf{0}}\rangle$. The 
total reflection is $N_{\textbf{0}}=\langle{\textbf{0}}\lvert{\textbf{0}}\rangle+\langle{\textbf{0}}\lvert{\textbf{1}}\rangle$. 
The reflected capital must carry a phase factor $e^{i\pi}$,
\begin{eqnarray}
N_{\textbf{0}}=e^{i\pi}[\langle{\textbf{0}}\lvert{\textbf{0}}\rangle+\langle{\textbf{0}}\lvert{\textbf{1}}\rangle].
\end{eqnarray}
A winning game requires more transmitted capitals than the reflected ones. 
The final gain of the game, $G(t)$, is 
\begin{equation}
G(t)=N_{\textbf{1}}+N_{\textbf{0}}>0.
\end{equation}
In the meantime, the sum of the two absolute values are the total number, 
$N={\lvert}N_{\textbf{1}}{\lvert}+{\lvert}N_{\textbf{0}}{\lvert}$. We can define normalized transmission 
$p={\lvert}N_{\textbf{1}}{\lvert}/N$ and normalized reflection $1-p={\lvert}N_{\textbf{2}}{\lvert}/N$ to 
get rid of the dependence on the absolute value of input. If the gain $G(t)<0$, the game loses.

Any game has an input and output. The output is either win or loss. If we input a positive capital, 
a positive output corresponds to the win, a negative output corresponds to the lose. If we add a negative sign 
on the input, the output will flip a sign correspondingly. If the game either 
transform the positive input capital into a negative output or 
transform the negative input into positive output, this is always a losing game, we call the game a flipper. 
A flipper check the sign of input and sent out an output with opposite sign. The simplest flipper is a mirror with 
total reflection.

Single game consisting of odd number of flipper is losing game, while the combination of even number of flipper is 
a winning game. For example, we take a sequence of flippers, $\{A,B,C,D,\cdots\}$ (Fig. \ref{convert} (b)). The flipper 
A maps the input capital $C_{in}$ to $C_{out}$ under the constraint $C_{in}{\times}C_{out}=-1$, so do the flippers 
$\{B,C,D,\cdots\}$. Different flipper produce different ratio between output capital and input capital, we denote 
this ratio $C_{out}/C_{in}$ as amplification, $F_{\alpha}=C^{\alpha}_{out}/C^{\alpha}_{in}$, here $\{\alpha=A,B, C,\cdots\}$ 
is the label of the flipper. The amplification of a flipper is always negative, $F_{\alpha}<0$. If we play flipper A, it always lose, any positive capital turns out to be negative. 
But flipper A also transform negative capital into positive capital. If we play two flipper $A$, the output of the first A 
is the input of the second A, then the positive input capital at the first A will give out a positive output at the second A. 
The amplification of the double A game is $F_{AA}=F_{A}F_{A}>0$, $F_{A}<0$. If we play a sequence of flipper
$\{A,B,C\}$, the final amplification is $F_{ABC}=F_{A}F_{B}F_{C}<0$. For the most general case, the product of a sequence of 
flipper, $\{\alpha{i},i=1,2,3,\cdots\}$, the final amplification is 
\begin{eqnarray}
F=\prod_{{i}=1}^{n}F_{\alpha{i}}.
\end{eqnarray}
If $n=2k+1$, $F<0$, it is losing game. If $n=2k$, $F>0$, it is winning game.

The input and output of any flipper must be classical variables. The player is actually a classical observer. The player
 make decisions following classic logic. A flipper may have complex internal game process dealing with 
complex quantum states. When we try to check the output of a flipper to see it wins or loses, 
we must measure the weight of those states, all the results will be summarized into a classical number. Therefore the 
inter-junction between independent flippers is always classical operation, no quantum states could survival. 
Quantum states may be used to design the internal gaming process within a flipper. If two independent flipper are 
perfectly organized into one combined flipper, then the old interface between single flippers disappeared. 
The quantum states can flow through those old classical channels. If we cut the united flipper into two parts without measuring 
the cutting edge, the quantum states are still waiting on the interface. But we can not determine wether each part wins or lose. 
They will collapse into classical numbers if we perform measurement.

\begin{figure}
\begin{center}
\includegraphics[width=0.35\textwidth]{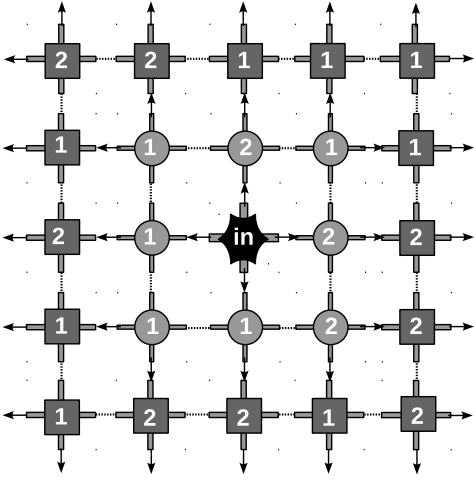}
\caption{\label{many}The dark star in the center is where the game begins, input capital is injected there. The box with label $\textit{1}$ 
represents the direct product of odd number of elementary flippers. The box labeled by $\textit{2}$ represents the direct combination of 
even number of flippers. The cups attached to the wall represent quantum state collector.}
\end{center}
\vspace{-0.5cm}
\end{figure}

Strictly speaking, there is no quantum game paradox as long as we want a definite output state of either winning or losing 
for each sub-flipper. Quantum states can be used to design the internal structure of single flipper. For example, 
we choose electron with spin as player. The winning state is more spin-up than spin down, the losing state is the opposite case. 
The input electrons may be all spin-ups. Then we design various complex scattering process within the flipper, 
such as two slits, many slits, magnetic field, electric field, other fixed nuclear spin, and so on. On the output 
side, we measure the state of all output electron. If the total number of spin-ups is more than those spin down, the game wins, 
this is not successful flipper. This design must be abandoned and start a new design until we get more spin-down than spin-up.    
Once we get a flipper with quantum internal structure, and make a good interface between the output of one 
flipper and the input another flipper, one can play the game paradox using many copies of the same flipper. From a global
 point of view, this is a classical game. If we go inside every flipper, it is a quantum processor.

A single flipper does not results in so-called game paradox. It is only when we combine many flippers in a complex way, 
the player can not read out the final output at first sight, the total game will produce seemly strange results. 
To find out the output of flipper network has a lot of things in common with many body physics. Especially when the 
flipper are quantum processors, we are essentially manipulating various entangled quantum states from local flipper 
to reach a global winning output. I showed an example of combinate game on $5\times5$ square lattice(Fig. \ref{many}). 
This game is the combination of 25 sub-games. There are 14 losing games, and 
11 winning games. The winning game are performed by the winning processor which is produced by even number of 
elementary flippers. They are represented by the boxes labeled by $\textit{2}$. The processor produced by odd number of elementary flipper 
are in charge of the losing games, they are labeled by $\textit{1}$(Fig. \ref{many}). 
The game starts from the dark star in the center. The dark star is a losing processor, it gives out four outputs. At the next step, 
the four outputs flow into a ring of eight processors. Three of them wins(the disc with label $\textit{2}$ in Fig. \ref{many}) 
The other five would lose(the disc with label $\textit{1}$ in Fig. \ref{many}). The $12$ outputs of this ring 
provide the input for the last round of game. There are $16$ processors in the last round, $8$ lose and 8 win. The output result at the first round of game can be read out by performing measurement on each of the 12 output. 
The output states from winning processors are projected to final winning state 
$\langle\textbf{1}\lvert$, this give the positive gain. While the output of those losing processors counts the 
negative gain. The total gain is the sum of all the outputs of this round, 
\begin{eqnarray}
G(t)=\sum_{i,j,\alpha}\left[\langle\textbf{1}_d\lvert\psi_{_{d(i,j)}}^{\alpha}\rangle
-\langle\textbf{0}_d\lvert\psi_{_{d(i,j)}}^{\alpha}\rangle\right],
\end{eqnarray}
the indexes $i,j$ runs over all the eight processors represented by disc in Fig. \ref{many}. If $G(t)>0$, the game wins, 
otherwise the games loses. For a game combined $n\times{n}$ subgames, the total number of output at the last round of game is 
$n^2+4$. If every sub-game is different from another, the dimension of probability variables is $n^2$. 
However we have only one equation to solve, $G(t)>0$. Even if all the sub-games are losing, the solution space is a 
$n^2$ dimensional cubic, each edge of the cubic has a length of $[0,1/2)$. 
To solve one inequality equation of $n^2$ unknown variables is a trivial problem. But if all the subgames 
labeled by $\textit{1}$ are the same, so does the game with label $\textit{2}$. We have only two variables which are distributed 
on the whole lattice, it is an interesting question to ask wether the combined game wins or loses.

To find out the final output of flipper network has a lot of things in common with many body physics. Especially when the 
flipper are quantum processors, we are essentially manipulating various entangled quantum states from local flipper 
to reach a global winning output. Many coupled identical flippers on a regular lattice can be theoretically summarized into a many body Hamiltonian system. 
Each flipper represents one losing game, this many body system represent a complex game of many coupled losing sub-games. 
We interested in the existence of a winning combined game 
out of many losing games, or the opposite case. Although this is no longer a paradox 
due to the strong coupling between neighboring losing games, 
it is still an interesting question. I will take many Ising spins on square lattice 
as an example to show how to model a quantum many body system into 
a combined game of many losing sub-games. We use Pauli matrix 
$\sigma^{x}_{i}$ to express a flipper. Each flipper occupied one lattice site. The nearest 
neighboring flippers couple with each other to work together. 
The output of the flipper pair $\sigma^{x}_{i}\sigma^{x}_{j}$ for the input state 
$\lvert{s_{i}s_{j}}\rangle$ is 
\begin{eqnarray}
h_{ij}=\langle{s_{i}s_{j}}\rvert\sigma^{x}_{i}\sigma^{x}_{j}\lvert{s_{i}s_{j}}\rangle,
\end{eqnarray}
$s_{i}=\pm1$ is Ising spin. The output operator of the combined game on the whole lattice is Hamiltonian
\begin{eqnarray}\label{ising}
H=\sum_{<ij>}J_{ij}\sigma^{x}_{i}\sigma^{x}_{j},
\end{eqnarray}
where $<ij>$ denotes the nearest neighboring lattice sites. $J_{ij}$ is the coupling strength. The input capital is 
the difference between the total 
number of spin $+1$ and the total number of spin down $-1$ in any given initial 
state  
\begin{eqnarray}
\lvert{\psi_{input}}\rangle=\lvert{s_{1}s_{2}s_{3}\cdots s_{n}}\rangle.
\end{eqnarray}
For instance, $\lvert{\psi_{input}}\rangle=\lvert{1,-1,1,\cdots,-1,-1,1,}\rangle$. The game rule is to minimize the 
total energy to reach a ground state. Playing one round of game is performed by operating the Hamiltonian 
on the initial configuration once. The output state after $k$ rounds of game is 
\begin{eqnarray}
\lvert{\psi^k_{output}}\rangle=H^k\lvert{\psi_{input}}\rangle.
\end{eqnarray}
The spin configuration at each round of game must make sure that the total energy 
$E=\langle{\psi}\rvert{H}\lvert{\psi}\rangle$ decreases comparing with last round of game. 
We take $s_{i}=+1$ as the positive capital for the game, if it becomes $-1$ in the end, the game lost. The 
total game covers many sub-games. If we input less 
$\{s_{i}=+1\}$, but end up with more $\{s_{i}=+1\}$ than our input, the games wins. 
The Hamiltonian Eq. (\ref{ising}) is 
equivalent to the familiar Ising Hamiltonian $H=\sum_{<ij>}J_{ij}\sigma^{z}_{i}\sigma^{z}_{j}$.
Theoretical physicist always search for the stable spin configurations of ground 
state starting from an arbitrary initial configuration, especially in quantum monte Carlo simulations. The so-called 
game paradox appears almost everywhere in many different quantum models. In mind of the great interest on 
time series in stochastic dynamics, 
what quantum Monte Carlo simulation has abandoned during the process of finding ground states 
maybe can help people to design winning games by combining a large number of lost sub-games. On the other hand, 
the so-called game paradox remind us that designing different non-equilibrium game process maybe leads to different 
physical properties of many body system.

\section{Summary}

When we divide a complex winning game into many sections, each section has an input and output, we call it a sub-game. 
Every sub-game must be tested to see wether it is a losing or winning game. It is not against intuition 
that if all sub-games are lost. More over, if we combine these sub-games together 
but in a different order as the original one, it is not against intuition either if 
the combined game is a losing game. No mater it is dividing or combining, we 
are dealing with the correlations between those sub-games. A real game paradox 
only exist when there is no any correlation between sub-games. In that case, any conclusion about 
the total game based on the result of sub-games are not reasonable.

Both Parrondo-Harmer-Abbott game and Parrondo's game introduced the correlation of probabilities between the nearest 
two rounds of game. In mind of a fact that the first round and the second round are the nearest neighbor, 
the second round and the third round are also nearest neighbor, and so forth, it forms a long chain of many coupled sub-games 
in history. 
In my optical analogy, the photon propagates through a long path connecting many beam splitters. The photon 
is either reflected or transmitted at each beam splitter. 
The final intensity of this photon is the product of the reflection or transmission coefficient of every beam splitter 
along this path. The more rounds of game one played, the more paths one would have. We sum up the intensity of 
all possible paths to reach the final output. If we have more transmitted photon than reflected photons, the final gain 
would be positive, the game wins, otherwise the game lost. If the photon is reflected for odd number of times, 
its contribution to the final gain is negative. If the photon is reflected even number of times, its contribution is 
positive. Since reflection photon can transmit through or reflected by other beam splitter, the lost of one 
local sub-game can become positive when a negative photon is reflected again by another beam splitter. 
Both Parrondo-Harmer-Abbott game and Parrondo's game can be mapped exactly into 
the propagation of a photon through the array of beam splitters.

To implement Parrondo-Harmer-Abbott game, we 
can use the same type of physical beam splitters with different transmission and reflection coefficient. 
While to implement Parrondo's old game, we need two different types of beam splitter: one add a $e^{i\pi}$ phase 
shift to its reflection, the other does not. There is modular operation in Parrondo's old game that 
the capital in game must be checked to see if it can be divided by a integer or not. This modular operation 
is equivalent to a special beam splitter from the point view of probability distribution. If we put this special 
beam splitter aside and only compare the optical diagram structure of Parrondo-Harmer-Abbott game and 
Parrondo's old game, they are essentially same. Parrondo-Harmer-Abbott game and Parrondo's game simulate 
the random sequence of combing game A and game B, the final game is winning. In fact, what really matters is 
the coupling between the two steps in game B, no matter how random the sequence of $ABAABBA\cdots$ are, it does not 
break the internal coupling inside game B. It is through the coupled two steps, or in other words, through two 
neighboring beam splitters, the probability of negative capital is transformed into the probability of 
positive capital. This lies in the heart of a winning combined game out of losing games.

We come to the conclusion: there is no paradox in both Parrondo-Harmer-Abbott game and Parrondo's old game, 
these two games are both reasonable and share the same strategy of coupling the nearest two rounds of game. 
This is a good news to gamblers for they can win from 
loss by reasonable calculations without worrying about paradox.

One can design much more complex games by drawing optical diagrams following the same strategy I used. A three-step 
history-dependent game is designed as an example to show how to combine many games. In this game, even if all 
the sub-games lost, the output of this combined game can win, lose or oscillating between loss and win. Different 
probability distribution leads to different results. To calculate the final output, one can draw 
all the paths connecting beam splitters. The final gain of each path is derived by 
multiplying the value of probability of all the bonds along this path. Summing up all the paths gives 
the final gain. A real photon propagating across array of beam splitters provide a practical way 
to test the complex design.

Since correlation can make losing games win, it is a good strategy to implant more correlations among sub-games.  
I designed a history entangled game with only two sub-games. The input state of each sub-game depends on the symmetric or 
antisymmetric combination of the historical states two steps earlier. In the symmetric states, two winning sub-games may leads to 
a lost combined game. If one of them wins, the game can win. In the antisymmetric state, two losing game results in 
a winning combined game. If one wins, the other lost, the output is either lose or win, it depends on specific parameters. 
In some parameter region, the final gain oscillates between win and loss.

We can implant strong correlations into the capital instead of sub-games. This is in case if one can 
not modify the game rule, but one can change the way of playing the capital in hand. Suppose one has $1000$ points 
in the beginning, we cut it into many pieces, $100+350+10+50+\cdots=1000$. Then we invest the capital in game according 
to a wave pattern, such as $\cos[t]^2+\sin[t+\tau]^2$, $\tau$ is the delay time. When this capital wave propagates 
through the lattice of many coupled sub-games, the outputs of sub-games are strongly entangled. To determine 
the final output of these quantum entangled states is equivalent to solve a quantum many body physics problem. 
For example, when we search for the spin configuration of ground state on a lattice, usually it starts from a 
random initial spin configuration, this initial configuration can be viewed as input capital of a game. Then we 
update the initial configuration by applying a functional of Hamiltonian operator, this is a process of combined game, 
every local operator on single lattice is a sub-game. The final output of the game is stable spin configuration. 
This is how quantum Monte Carlo simulation runs everyday. In fact, Monte Carlo method, 
as its name suggested, comes from gamble games.

Coupled many sub-games is useful for designing various complex Brownian ratchet. The one dimensional periodic asymmetric potential 
is equivalent to a chain of many beam splitters. If a flashing Brownian can cover two or more potential well within 
one period, we must take into account of the coupling between two or more neighboring potential wells. The forward velocity is 
determined by summing up continuous two steps of pass or refelection. The backward velocity is determined by the joint probability of 
one reflection followed one pass or vice verse. The optical 
model provide a tool to analyze the relationship between different individual potential wells. 
Especially when there are many different individual local potential wells distributed along the long chain. 
If each local potential well has different reflection and transmission, it is hard to see 
the physics directly from Fokker-Planck equation or Smoluchowski equation. In that case, we take every local potential well 
as a different beam splitter, draw the optical diagram, then one will get at least a phenomenological understanding.

\section{Acknowledgment}

\end{document}